# An ordinary state-based peridynamic elastoplastic 2D model consistent with J2 plasticity

Farzaneh Mousavi, Siavash Jafarzadeh, Florin Bobaru[*]

Department of Mechanical and Materials Engineering
University of Nebraska-Lincoln, Lincoln, NE 68588-0526 USA

**Abstract:**

We present a new ordinary state-based peridynamic model (OSB-PD) in 2D consistent with J2 plasticity using a novel decomposition for force and extension states. A new strategy for testing the consistency of an OSB-PD formulation for elastoplasticity is introduced. In contrast with other similar models, the new elasto-plastic OSB-PD model is objective and works for large rotations. We formulate two rate-independent yield functions equivalent to J2 plasticity with associated flow rules, based on force states and on the strain energy density. An efficient return mapping algorithm for peridynamic formulations in plasticity is proposed. We verify the model against results obtained with Abaqus for the classical formulation using two examples, one featuring plastic deformations under large rotations. The evolution of the plastic zone computed with the new model matches that from the classical model found with Abaqus.

## 1. Introduction

*Significance and motivation*

Peridynamic modeling of elastoplastic deformation and failure promises to deliver predictive results for some challenging problems related to ductile failure (Haltom et al., 2013). The correspondence non-ordinary state-based PD models (Behzadinasab et al., 2019), are limited to the types of material behavior that classical constitutive models can represent, and may also require certain corrections to eliminate material instabilities noticed in such formulations (Behzadinasab and Foster, 2020; Gu et al., 2018; Littlewood, 2012; Silling, 2017; Tupek and Radovitzky, 2014). Ordinary-state based peridynamic models (OSB-PD) are more general and do not suffer from instabilities. For elastoplastic behavior, OSB-PD models have been introduced in (Madenci and Oterkus, 2016; Mitchell, 2011; Silling et al., 2007). Several 2D versions of this theory appeared in (Madenci and Oterkus, 2016; Pashazad and Kharazi, 2019; Zhou et al., 2018).

---

[*] corresponding author
Email address: fbobaru2@unl.edu

In this paper we find some inconsistencies in the existing 2D OSB-PD models of elastoplasticity and we introduce a new formulation that is consistent with the stress decomposition in the classical J2 theory of plasticity (Simo and Hughes, 2006). In contrast with some existing 2D OSB-PD models, the new elastoplastic model, with plane stress/strain formulations, is objective and works for large rotations. The new decomposition of force and extension PD states can be used for other OSB-PD models. A reliable PD elastoplastic model, able to correctly predict plastic deformation, is a necessary step for future models of ductile failure and fracture.

*Literature review*

Peridynamics, a nonlocal alternative formulation of the classical continuum mechanics, does not use spatial derivatives. While in classical mechanics it is assumed that material point interactions are due to their mutual direct contact only, in peridynamics each point interacts with all other points within a finite distance called the horizon, $\delta$, via peridynamic "bonds". This approach was first introduced in (Silling, 2000) for elastic bodies and brittle fracture. Peridynamics can model evolving discontinuities like cracks in a domain, since it replaces spatial derivatives from the equations of motion with an integral operator. A variety of peridynamic models have been developed for: brittle fracture (Ha and Bobaru, 2010), thermally-driven cracks (Xu et al., 2018), failure in composites (Mehrmashhadi et al., 2019) and in porous media (Chen et al., 2019), and corrosion damage (Jafarzadeh et al., 2019b, 2019a, 2018). The original form of peridynamic model is "bond-based" theory (Silling, 2000). In bond-based peridynamics (BB-PD) the force density transmitted in each bond depends on the deformation of that bond only. This results in a fixed Poisson's ratio for BB-PD models. In addition, bond-based approaches cannot exactly enforce incompressible shear deformation, which is required in plastic deformations. To overcome such limitations, a more general framework , called "state-based" peridynamics was proposed (Bobaru et al., 2016; Silling et al., 2007). In state-based peridynamics, the bond force density between two points depends on the deformation of all other bonds connecting to the bond's end points. State-based peridynamics can be divided into ordinary and non-ordinary methods. In ordinary state-based peridynamics (OSB-PD) the bond force densities are aligned with the deformed bond vectors. In non-ordinary state-based peridynamics (NOSB-PD) however, the bond force is not restricted to be parallel to the deformed bond (Silling et al., 2007).

Correspondence models are a class of NOSB-PD, where the classical constitutive material models are adopted in the PD nonlocal settings (Silling et al., 2007). Most peridynamic models for simulating plastic deformations are of such correspondence type (Amani et al., 2016; Foster et al., 2010; Pathrikar et al., 2019; Rahaman et al., 2017; Sun and Sundararaghavan, 2014; Warren et al., 2009), since such models allow one to directly employ existing classical constitutive plasticity models into the PD framework (Silling et al., 2007). NOSB-PD

correspondence approaches, however, do have some deficiencies, mentioned above. The instabilities present in the NOSB-PD correspondence approach are due to zero-energy deformation modes of the family of nodes (Silling, 2017). Furthermore, correspondence modeling leads to some loss of information when the deformation state is reduced to a deformation gradient tensor (Silling and Lehoucq, 2010). In contrast, native PD constitutive modeling allows to fully exploit the advantages of nonlocality, since bond-level information is maintained all along. The first native (OSB-PD) constitutive model for 3D elastic-perfectly plastic materials was proposed in (Silling et al., 2007). Soon after, (Mitchell, 2011) utilized the constitutive equation of elastoplastic materials in (Silling et al., 2007) and proposed a nonlocal yield criterion in terms of 3D force states equivalent to J2 plasticity. (Lammi and Vogler, 2014) formulated a 3D pressure-dependent plasticity model in OSB-PD that considered inelasticity and failure of concrete as a quasi- brittle heterogeneous material. The Drucker-Prager model was used to relate PD variables to those in the classical theory.

Two-dimensional elastoplastic models, when applicable, offer significant computational cost reductions compared to full 3D models. Several works have introduced 2D state-based PD models for elastoplastic behavior. In (Madenci and Oterkus, 2016), an OSB peridynamic plasticity constitutive relations based on the von Mises yield criterion with linear isotropic hardening was presented for 1D, 2D and 3D cases. The model in (Madenci and Oterkus, 2016), is limited, however, to a specific influence function and it is not mentioned whether the obtained 2D formulation is for plane stress or plane strain conditions. This model was later extended to include hardening by (Pashazad and Kharazi, 2019). Another limitation of the models in (Madenci and Oterkus, 2016; Pashazad and Kharazi, 2019) is that they are not valid under arbitrary rigid-body rotations (Le and Bobaru, 2018; Madenci, 2017). Estimation of the plastic zone near the crack tips in rocks was studied in (Zhou et al., 2018) using a 2D OSB-PD model. The 2D formulation in (Zhou et al., 2018), adopts the 3D formulation suggested in (Mitchell et al., 2015). However, as we discuss in detail in section 3, the approach in (Zhou et al., 2018) does not lead to a correct two-dimensional decomposition for the extension and force states, since the stress tensors corresponding to their decomposed force states are not consistent with the classical decomposition of stress which would lead to an incorrect yield function. Moreover, it is not clear if the 2D formulation in (Mitchell et al., 2015) is for plane stress or plane strain conditions.

*Organization of the paper*

This paper is organized as follows: a brief overview of the OSB-PD model and decomposition of 2D states is given in Section 2; the new decompositions for force and extension states in 2D are presented in Section 3; derivation of the new elastoplastic model for 2D OSB-PD is shown in Section 4, where we also provide the corresponding return-mapping algorithm; Section 5 presents the numerical approach and the discretization scheme; numerical results for two elastoplastic problems, one undergoing large rotations, are shown in Section 6, where the PD results are compared with those obtained from Abaqus; conclusions are gathered in Section 7.

## 2. Brief review of ordinary state-based peridynamics

In peridynamics, each material point interacts with its neighboring points located inside a finite size neighborhood referred to as the horizon region. The horizon region $\mathcal{H}_x$ of a generic point with the position vector $\boldsymbol{x}$, is usually taken to be a disk in 2D and a sphere in 3D centered at $\boldsymbol{x}$ and with radius of $\delta$, the "horizon size" (See Figure 1). Material points that are inside $\mathcal{H}_x$ are denoted by their position vector $\boldsymbol{x}'$ and referred to as family points of $\boldsymbol{x}$. Bonds are objects that carry the pairwise information between $\boldsymbol{x}$ and $\boldsymbol{x}'$. States are mathematical objects (nonlinear mappings, in general), defined at each point and carry the information about the nonlocal interaction of $\boldsymbol{x}$ with all its family members $\boldsymbol{x}'$. States are functions of bonds and return the information associated with each bond, $\boldsymbol{x}' - \boldsymbol{x}$. To simplify the notation, let $\boldsymbol{\xi} = \boldsymbol{x}' - \boldsymbol{x}$ represent the bond vector. Note that in this study *bold-face* letters are used to denote vectors, and *underlined characters* are used to denote PD states. States can be scalar-states or vector states depending on which type of quantity they return. Here, lower-case and upper-case states refer to scalar and vector states respectively. For example, $\underline{a}[\boldsymbol{x}, t]\langle\boldsymbol{\xi}\rangle$ and $\underline{A}[\boldsymbol{x}, t]\langle\boldsymbol{\xi}\rangle$ are scalar and vector states defined at point $\boldsymbol{x}$ and time $t$. For more details on peridynamic states, please see (Silling et al., 2007).

We first define the PD states which are used in this study:

$\underline{X}$ denotes the *bond vector state* that returns bond vector in the reference configuration:

$$\underline{X}\langle\boldsymbol{\xi}\rangle = \boldsymbol{\xi}. \tag{1}$$

*Bond scalar state* $\underline{x}$ is defined by:

$$\underline{x}\langle\boldsymbol{\xi}\rangle = |\underline{X}\langle\boldsymbol{\xi}\rangle|. \tag{2}$$

Let $\boldsymbol{u}(\boldsymbol{x}, t)$ denote the displacement of $\boldsymbol{x}$ at time $t$. Then $\boldsymbol{y} = \boldsymbol{x} + \boldsymbol{u}$ is the position of $\boldsymbol{x}$ in the deformed configuration (See Figure 1). Then the *deformation vector state* $\underline{Y}$ is:

$$\underline{Y}\langle\boldsymbol{\xi}\rangle = \boldsymbol{y}(\boldsymbol{x}', t) - \boldsymbol{y}(\boldsymbol{x}, t) = \boldsymbol{y}' - \boldsymbol{y} \tag{3}$$

and the *deformation scalar state* is:

$$\underline{y}\langle\boldsymbol{\xi}\rangle = |\underline{Y}\langle\boldsymbol{\xi}\rangle| \tag{4}$$

A useful scalar state frequently used is the *extension state*:

$$\underline{e} = \underline{y} - \underline{x}, \tag{5}$$

which returns the magnitude of bond elongation.

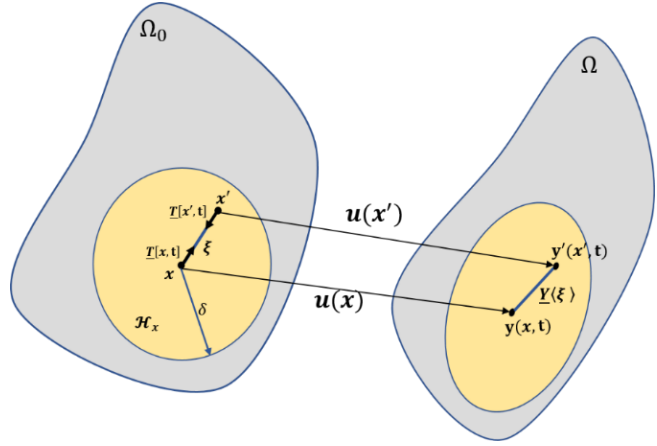

Figure 1: Schematic of deformations in a peridynamic body

In Figure 1, $\underline{T}[x, t]\langle\xi\rangle$, is the *force vector state* of $x$ which returns the force density associated with $x' - x$. Similarly, $\underline{T}[x', t]\langle-\xi\rangle$ is the force state at $x'$ which returns the force density associated with $x - x'$. In the ordinary state-based peridynamics the force vector state $\underline{T}[x, t]\langle\xi\rangle$ is parallel to the deformation vector state $\underline{Y}[x, t]\langle\xi\rangle$, in contrast with the non-ordinary PD model. The force vector state in an ordinary state-based peridynamics is defined as (Silling et al., 2007):

$$\underline{T} = \underline{t}\,\underline{M}, \tag{6}$$

where $\underline{t}$ is the *scalar force state* that gives the magnitude of the force vector state $\underline{T}$, and $\underline{M}$ is the unit vector state of deformation that returns the direction of the deformed bonds:

$$\underline{M} = \frac{\underline{Y}}{|\underline{Y}|} \tag{7}$$

The equation of motion for material point $x$ in the state-based peridynamics is (Silling et al., 2007):

$$\rho(x)\ddot{u}(x,t) = \int_{\mathcal{H}_x} (\underline{T}[x,t]\langle x'-x\rangle - \underline{T}[x',t]\langle x-x'\rangle)\mathrm{d}V_{x'} + b(x,t) \tag{8}$$

where $\rho$ is the mass density, $\ddot{u}(x, t)$ is the acceleration of the material point $x$ at time $t$, and $b$ is the body force per unit volume. $\underline{T}[x, t]\langle x'-x\rangle - \underline{T}[x', t]\langle x-x'\rangle$ determines the total force point $x'$ exerts on $x$.

## 2.1. Two-dimensional ordinary state-based peridynamic for isotropic linear elastic model

The original formulation for two-dimensional (2D) state-based peridynamic elasticity, which is used in this study, was introduced in (Le et al., 2014). The derivation of the 2D constitutive equation in (Le et al., 2014) follows the procedure introduced in (Silling et al., 2007) for 3D, modified for the 2D case: 1) given the classical strain energy density for a homogeneous deformation, the corresponding PD strain energy density is set to match it; this determines the unknown coefficients in the scalar extension state in terms of classical elasticity constants; and 2) taking the Fréchet derivative of the 2D peridynamic strain energy density with respect to extension state, delivers the 2D force state in terms of the extension state and elasticity constants. A model derived with this approach is called the Linear Peridynamic Solid (LPS) model (Silling et al., 2007), because of its similarity to a linear elastic solid in the classical continuum framework.

Let $\underline{a}$ and $\underline{b}$ be two generic scalar states. The inner product operation is defined as (Silling and Lehoucq, 2010):

$$\underline{a} \bullet \underline{b} = \int_{\mathcal{H}_x} \underline{a}\langle x' - x \rangle \underline{b}\langle x' - x \rangle \mathrm{d}V_{x'} \tag{9}$$

Given $k$, $\mu$ and $\nu$ the bulk modulus, shear modulus and Poisson ratio, respectively, the 2D peridynamic strain energy density derived in (Le et al., 2014) is:

$$W(\theta, \underline{e}) = \frac{k'\theta^2}{2} + \frac{\alpha}{2} \underline{\omega} \left( \underline{e} - \frac{\theta \underline{x}}{3} \right) \bullet \left( \underline{e} - \frac{\theta \underline{x}}{3} \right) \quad ; \quad \alpha = \frac{8\mu}{m} \tag{10}$$

In Eq.(10), $\underline{\omega}$ is the *influence function state* that modulates the nonlocal interactions (weighting the contribution of each bond, (Seleson and Parks, 2011)) and can be defined with respect to the bond length (e.g. $\underline{\omega}\langle \xi \rangle = 1, \frac{1}{|\xi|}, etc.$), $m = \underline{\omega}\underline{x} \bullet \underline{x}$, while $k'$ is defined as follows:

$$k' = \begin{cases} \frac{(7-11\nu)k}{6(1-2\nu)} & \text{plane stress} \\ k + \frac{\mu}{9} & \text{plane strain} \end{cases} \tag{11}$$

The scalar-valued function $\theta$ denotes the nonlocal volume dilatation under the assumption of small deformations:

$$\theta = \begin{cases} \frac{2(2\nu-1)}{(\nu-1)m} \underline{\omega}\underline{x} \bullet \underline{e} & \text{plane stress} \\ \frac{2}{m} \underline{\omega}\underline{x} \bullet \underline{e} & \text{plane strain} \end{cases} \tag{12}$$

The simplified version of force state for 2D plane stress/strain condition obtained in (Le and Bobaru, 2018) is:

$$\underline{t} = \begin{cases} \left(k - \frac{8\mu}{3}\right)\theta \frac{\underline{\omega}\underline{x}}{m} + \frac{8\mu}{m} \underline{\omega}\underline{e} & \text{plane stress} \\ 2\left(k - \frac{5\mu}{3}\right)\theta \frac{\underline{\omega}\underline{x}}{m} + \frac{8\mu}{m} \underline{\omega}\underline{e} & \text{plane strain} \end{cases} \tag{13}$$

The detailed derivation of the PD strain energy density, the force state and scalar function $\theta$ (volume dilatation) can be found in (Le et al., 2014).

## 3. Decomposition of 2D peridynamic states into deviatoric and isotropic parts

Decomposing the extension state and force state into co-isotropic and co-deviatoric parts is a significant step in describing elastoplastic material response because plastic deformation only results from shear. Therefore, having the correct deviatoric force and extension states is critical. As we show below, the decompositions available in the literature for a formulation applicable to large rotations are not consistent with the classical theory, in the sense that the stress tensors obtained via the correspondence procedure (finding the isotropic /deviatoric part of stress from isotropic/deviatoric force state) are not equivalent to the classical decomposition of the stress tensor (Mitchell et al., 2015). PD plasticity models that rely on such formulations may be deficient. In this section, we introduce a new decomposition for force and extension states in 2D ordinary state-based peridynamics consistent with the deviatoric/hydrostatic decomposition of stress and strain tensors in the local theory.

### 3.1. A brief review of peridynamic isotropic and deviatoric stress tensors

In this section, we discuss the peridynamic *stress tensor* which is used in next section to find the decomposition of force states. The PD stress tensor is the flux of linear momentum through a surface (Silling and Lehoucq, 2008). It has been shown ((Silling and Lehoucq, 2008)) that, in the limit, the PD stress tensor approaches the collapsed PD stress tensor ($\sigma_{\text{PD}}$) which is equivalent to the first Piola-Kirchoff stress tensor. The expression for the PD collapsed stress tensor obtained from the corresponding force state is:

$$\sigma_{\text{PD}} = \int_{\mathcal{H}} \underline{T}\langle\boldsymbol{\xi}\rangle \otimes \boldsymbol{\xi} \, dV_{\xi} \tag{14}$$

As it is shown in Figure 2, given the force state $\underline{T}$, there are two ways to find the isotopic and deviatoric parts of the corresponding $\sigma_{\text{PD}}$:

(a) First convert the PD force state into the PD stress tensor via Eq. (14), then use the classical definition of stress tensor decomposition (see Appendix A) to obtain the isotropic and deviatoric parts of $\sigma_{\text{PD}}$;

(b) First decompose the force state into its isotropic and deviatoric parts ($\underline{T}^{\text{iso}}$ and $\underline{T}^{\text{d}}$), then use Eq. (14) to collapse each part to their corresponding stress tensors:

$$\sigma_{\text{PD}}^{\text{iso}} = \int_{\mathcal{H}} \underline{T}^{\text{iso}}\langle\boldsymbol{\xi}\rangle \otimes \boldsymbol{\xi} \, dV_{\xi} \quad , \quad \underline{T}^{\text{iso}}(\underline{Y}) = \underline{t}^{\text{iso}}(\underline{Y})\underline{M}(\underline{Y}) \quad , \quad \underline{X}\langle\boldsymbol{\xi}\rangle = \boldsymbol{\xi} \tag{15}$$

$$\sigma_{\text{PD}}^{\text{d}} = \int_{\mathcal{H}} \underline{T}^{\text{d}}\langle \underline{\xi} \rangle \otimes \underline{\xi} \, dV_\xi \ , \quad \underline{T}^{\text{d}}(\underline{Y}) = \underline{t}^{\text{d}}(\underline{Y}) \underline{M}(\underline{Y}) \tag{16}$$

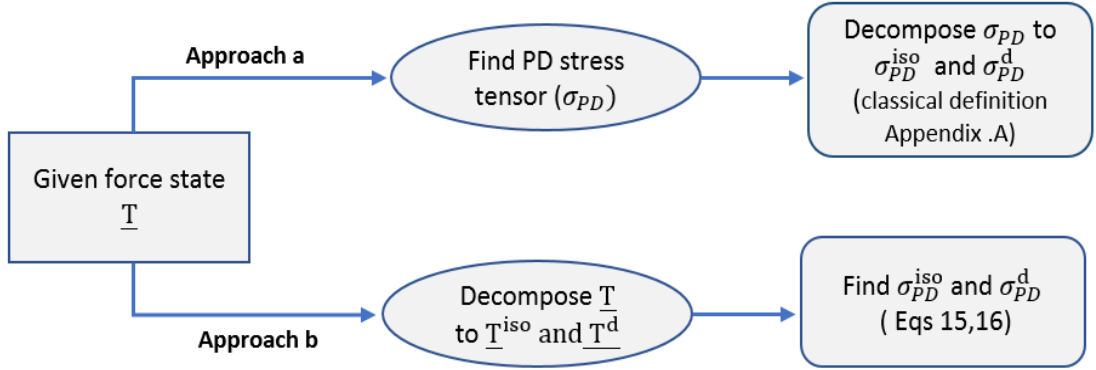

Figure 2 : Two equivalent approaches for obtaining the isotropic/deviatoric decomposition of the stress tensor corresponding to a given force state.

These two approaches should be equivalent (lead to the same tensors). Consequently, one way to verify whether a proposed decomposition of force states is self-consistent is to check whether the two approaches mentioned above indeed lead to the same isotropic and deviatoric stress tensors. Note that the obtained $\sigma_{PD}^{\text{iso}}$ and $\sigma_{PD}^{\text{d}}$ (from Eqs. (15) and (16)) need to satisfy the expected properties of deviatoric and isotropic tensors discussed in Appendix A. For instance, the first invariant of the deviatoric collapsed stress tensor has to be zero.

### 3.2. A new decomposition of peridynamic force and extension states in 2D

In this section, we introduce decompositions for the force and extension states for 2D plane stress and plane strain conditions that are consistent with the classical theory.

#### 3.2.1. Decomposing 2D force state into isotropic and deviatoric part

In this section, first we find the 2D isotropic force state ($\underline{t}^{\text{iso}}$) and check its self-consistency (according to what is explained in previous part). The deviatoric part of force state is defined by $\underline{t}^{\text{d}} = \underline{t} - \underline{t}^{\text{iso}}$. In order to define a 2D isotropic force state, we use Eq. (15) which considers the correspondence relation between the isotropic force state and PD isotropic stress tensor. Here we use a different approach from the one used in (Silling et al., 2007) for the similar decomposition but performed in 3D. We have verified this new approach on the 3D case as well, and we recovered the formulas shown in (Silling et al., 2007).

Referring to Eq. (15), we try to define $\underline{t}^{\text{iso}}$ in a way which produces a PD isotropic stress tensor equivalent to the classical isotropic (hydrostatic) stress tensor. For this purpose, first we look at the general form of stress tensor and its isotropic part. Hooke's law for an isotropic linear material is:

$$\sigma_{ij} = \left(k - \frac{2}{3}\mu\right)\delta_{ij}\epsilon_{kk} + 2\mu\epsilon_{ij}; \quad i,j = 1,2,3; \quad \epsilon_{kk} = \epsilon_{11} + \epsilon_{22} + \epsilon_{33} = \epsilon_V \tag{17}$$

where $\epsilon_{ij}$ is the $ij$ component of the infinitesimal strain tensor and $\epsilon_{kk}$ is the volume dilatation. According to the definition of hydrostatic (isotropic) stress tensor, we have:

$$\sigma^{iso} = \frac{1}{3}\sigma_{kk}\delta_{ij} = \frac{1}{3}\left[\left(k - \frac{2}{3}\mu\right)\delta_{kk}\epsilon_{kk} + 2\mu\epsilon_{kk}\right]\delta_{ij} = \frac{1}{3}(3k\,\epsilon_{kk})\delta_{ij} = k\epsilon_{kk}\delta_{ij} \tag{18}$$

It is important to emphasize that the definitions above are identical for 3D or 2D plane stress/strain models. Note that, collapsing a 2D peridynamic state into a stress tensor using Eq. (14), results in a 2D tensor, meaning that the out-of-plane stress components cannot be obtained in this fashion. The out-of-plane stress components ($\sigma_{33}$, $\sigma_{33}^{iso}$, $\sigma_{33}^{d}$) should be calculated after the state-to-tensor conversion by using known relationships in plane stress/strain conditions if needed (Boresi et al., 1985). Considering Eq. (18) for such 2D tensors we have:

$$\sigma_{11}^{iso} + \sigma_{22}^{iso} = 2k\epsilon_{kk}. \tag{19}$$

Now we use Eq. (15) to find $\underline{t}^{iso}$ formula. Eq. (15), in indicial notation is:

$$\sigma_{ij}^{iso} = \int_{\mathcal{H}} \underline{T}_i^{iso}\langle\xi\rangle \underline{X}_j\langle\xi\rangle \, dV_\xi \quad , \quad \underline{T}^{iso} = \underline{t}^{iso}\underline{M} \tag{20}$$

$$\sigma_{ij}^{iso} = \int_{\mathcal{H}} \underline{t}^{iso}\langle\xi\rangle \underline{M}_i\langle\xi\rangle \underline{X}_j\langle\xi\rangle \, dV_\xi \tag{21}$$

where $\underline{M}$ is a unit vector state in the deformed bond direction, and $i,j = 1,2$ in 2D. With the assumption of small deformations, the deformed configuration is the same as the undeformed configuration. Observing the notations in Figure 3 we can write:

$$\sigma_{ij}^{iso} = \int_{\mathcal{H}} \underline{t}^{iso}\langle\xi\rangle (\underline{M}\langle\xi\rangle \cdot \hat{e}_i) \, (\underline{X}\langle\xi\rangle \cdot \hat{e}_j) \, dV_\xi \tag{22}$$

Using Eq.(22) in Eq. (19), we get:

$$\int_{\mathcal{H}} \underline{t}^{iso}\langle\xi\rangle \underline{x}\langle\xi\rangle \cos\phi \, \cos\phi \, dV_\xi + \int_{\mathcal{H}} \underline{t}^{iso}\langle\xi\rangle \underline{x}\langle\xi\rangle \sin\phi \, \sin\phi \, dV_\xi = 2k\epsilon_{kk} \tag{23}$$

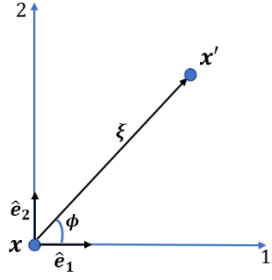

Figure 3: Representation of a bond in 2D coordinate system

Similar to the relation for 3D $\underline{t}^{iso}$ suggested in (Silling et al., 2007), we assume that the scalar state $\underline{t}^{iso}$ has the following form:

$$\underline{t}^{iso} = \beta \, \underline{\omega x} \tag{24}$$

where $\beta$ is a scalar to be determined. Substituting $\underline{t}^{iso}$ from Eq. (24) into Eq. (23), results in:

$$2k\epsilon_{kk} = \beta \int_{\mathcal{H}} \underline{\omega x}\langle\xi\rangle \, \underline{x}\langle\xi\rangle dV_\xi = \beta m \tag{25}$$

Accordingly:

$$\beta = \frac{2k}{m}\epsilon_{kk} \tag{26}$$

We substitute Eq. (26) into Eq. (24) and change $\epsilon_{kk}$ to $\theta$, since the peridynamic volume dilatation $\theta$ is defined to be equal to $\epsilon_{kk}$ for small strains. The formula for $\underline{t}^{iso}$ in 2D state-based peridynamics is then determined as:

$$\underline{t}^{iso} = \frac{2k}{m}\theta \, \underline{\omega x} \tag{27}$$

Note that, $\theta$ for 2D plane stress/strain conditions is given by Eq. (12). The decomposition presented here is different from the 2D decomposition in the OSB-PD formulation presented in (Mitchell et al., 2015). The 2D PD formulation in (Mitchell et al., 2015) is a "pure" 2D model in which the definition of the isotropic part of the force state does not include the third direction information present in 2D plane stress/strain modes. Therefore, such a model would not be applicable to plane stress/strain problems in plasticity. Another decomposition has been obtained in (Madenci and Oterkus, 2016), where $\theta$'s definition is valid for no or infinitesimal rotations only, and it was not specified whether the derivation was for plane stress or plane strain conditions. In (Le and Bobaru, 2018) it was shown that, in the case of using special influence functions, the 2D formula provided in (Madenci and Oterkus, 2016) is only valid for plane strain conditions.

Eq. (27) gives the isotropic part of the 2D force state in ordinary state-based peridynamics. Then, using Eq. (13), its deviatoric part is:

$$\underline{t}^d = \underline{t} - \underline{t}^{iso} = \begin{cases} \left(-k - \dfrac{8\mu}{3}\right)\theta\dfrac{\omega \underline{x}}{m} + \dfrac{8\mu}{m}\underline{\omega e} & \text{plane stress} \\ -\dfrac{10\mu}{3}\theta\dfrac{\omega \underline{x}}{m} + \dfrac{8\mu}{m}\underline{\omega e} & \text{plane strain} \end{cases} \tag{28}$$

**Remark:** Observe that the constitutive models in this study can be used for small deformations but arbitrary rotations. This is because the force state depends on the scalar bond extension ($\underline{e}$) state and the direction of the force vector state is the same as that of the deformed bond vector, this being an ordinary state-based model (which is critical for this property to hold!). Note also that the corresponding formulas in (Madenci and Oterkus, 2016) can only be applied to the case of no/small rotations because the volume dilation ($\theta$) function, as defined there, is not rotation independent (see the detailed discussion about this issue in (Le and Bobaru, 2018)).

As mentioned before, one way to check that the force state decomposition is consistent with the classical theory is obtaining the same stress tensors parts from the scheme shown in Figure 2 (see section 3.1). In the following, we show that the decomposition above satisfies this equivalency. For this, we compare $\sigma_{11}^{iso}$ obtained from the two approaches ("a" and "b" Figure 2):

Using approach "a" for plane stress conditions, we have:

$$\sigma_{11}^{iso(a)} = \frac{1}{3}[(\sigma_{PD})_{11} + (\sigma_{PD})_{22} + (\sigma_{PD})_{33}] \tag{29}$$

Substituting $\sigma_{PD}$ given in Eq. (14) in Eq. (29) gives:

$$\sigma_{11}^{iso(a)} = \frac{1}{3}\left(\int_{\mathcal{H}} \underline{T_1}\langle\xi\rangle \underline{X_1}\langle\xi\rangle \, dV_\xi + \int_{\mathcal{H}} \underline{T_2}\langle\xi\rangle \underline{X_2}\langle\xi\rangle \, dV_\xi + 0\right) \tag{30}$$

where $dV_\xi = l_z \underline{x}\langle\xi\rangle \, dx d\theta$, and $l_z$ is the thickness in z (out of plane) direction. We find:

$$\sigma_{11}^{iso(a)} = \frac{1}{3}\left(\int_{\mathcal{H}} \underline{t}\langle\xi\rangle \cos\phi \, \underline{x}\langle\xi\rangle \cos\phi \, dV_\xi + \int_{\mathcal{H}} \underline{t}\langle\xi\rangle \sin\phi \, \underline{x}\langle\xi\rangle \sin\phi \, dV_\xi\right) \tag{31}$$

$$= \frac{1}{3}\left(\int_0^{2\pi}\int_0^{\delta} \underline{t}\langle\xi\rangle \underline{x}^2\langle\xi\rangle \cos^2\phi \, l_z \, d\underline{x}\langle\xi\rangle \, d\phi\right.$$

$$\left. + \int_0^{2\pi}\int_0^{\delta} \underline{t}\langle\xi\rangle \underline{x}^2\langle\xi\rangle \sin^2\phi \, l_z \, d\underline{x}\langle\xi\rangle \, d\phi\right) = \frac{2\pi l_z}{3}\int_0^{\delta} \underline{t}\langle\xi\rangle \underline{x}^2\langle\xi\rangle \, d\underline{x}\langle\xi\rangle$$

Substituting $\underline{t}$ for plane stress conditions (see Eq. (13)) in Eq. (31) we have:

$$\sigma_{11}^{\text{iso}(a)} = \frac{2\pi l_z}{3} \left( \int_0^\delta k\theta \frac{\underline{\omega}\langle\xi\rangle}{m} \underline{x}^3\langle\xi\rangle \, d\underline{x}\langle\xi\rangle \right. \tag{32}$$

$$\left. - \int_0^\delta \frac{8\mu}{3} \theta \frac{\underline{\omega}\langle\xi\rangle}{m} \underline{x}^3\langle\xi\rangle \, d\underline{x}\langle\xi\rangle + \int_0^\delta \frac{8\mu}{m} \underline{\omega}\langle\xi\rangle \underline{x}^2\langle\xi\rangle \underline{e}\langle\xi\rangle \, d\underline{x}\langle\xi\rangle \right)$$

Note that $m = 2\pi l_z \int_0^\delta \underline{\omega}\langle\xi\rangle \underline{x}^3\langle\xi\rangle d\underline{x}\langle\xi\rangle$ and $\theta = 4\frac{2\nu-1}{\nu-1} \pi l_z \int_0^\delta \frac{\underline{\omega}\langle\xi\rangle}{m} \underline{x}^2\langle\xi\rangle \underline{e}\langle\xi\rangle \, d\underline{x}\langle\xi\rangle$ in 2D plane stress, hence Eq. (32) becomes:

$$\sigma_{11}^{\text{iso}(a)} = k\theta \tag{33}$$

Now we find $\sigma_{11}^{\text{iso}}$ based on approach "b", namely $\sigma_{11}^{\text{iso}(b)}$ (see Eq. (15)):

$$\sigma_{11}^{\text{iso}(b)} = \int_{\mathcal{H}} \underline{T}_1^{\text{iso}}\langle\xi\rangle \underline{X}_1\langle\xi\rangle \, dV_\xi \tag{34}$$

$$= \int_{\mathcal{H}} \underline{t}^{\text{iso}}\langle\xi\rangle \cos\phi \, \underline{x}\langle\xi\rangle \cos\phi \, dV_\xi$$

$$= \int_0^{2\pi} \int_0^\delta \underline{t}^{\text{iso}}\langle\xi\rangle \underline{x}^2\langle\xi\rangle \cos^2\phi \, l_z \, d\underline{x}\langle\xi\rangle \, d\phi$$

$$= \pi l_z \int_0^\delta \underline{t}^{\text{iso}}\langle\xi\rangle \underline{x}^2\langle\xi\rangle \, d\underline{x}\langle\xi\rangle$$

Using our decomposition for $\underline{t}^{\text{iso}}$ given in Eq. (27) and using $m = 2\pi l_z \int_0^\delta \underline{\omega}\langle\xi\rangle \underline{x}^3\langle\xi\rangle d\underline{x}\langle\xi\rangle$, we get:

$$\sigma_{11}^{\text{iso}(b)} = \pi l_z \int_0^\delta \frac{2k\theta}{m} \underline{\omega}\langle\xi\rangle \underline{x}^3\langle\xi\rangle \, d\underline{x}\langle\xi\rangle \tag{35}$$

$$= k\theta$$

As we see, both approaches give the same result for $\sigma_{11}^{\text{iso}}$. Repeating this type of calculations for the rest of the components, we get the $\sigma^{\text{iso}}$ stress tensor consistent with the classical theory, and therefore $\sigma^d = \sigma - \sigma^{\text{iso}}$ will be consistent with the classical theory as well. Similarly, one can easily show that results are consistent for plane strain conditions.

### 3.2.2. Decomposing 2D extension state into isotropic and deviatoric parts

Because the deviatoric part of extension state is responsible for the plastic deformation, it is necessary to have a consistent decomposition of the extension state for our elastoplastic model. To introduce a consistent decomposition of the 2D extension state, we rearrange Eq. (28) to express $\underline{e}$ in terms of $\underline{t}^d$ and $\theta$:

$$\underline{e} = \begin{cases} \dfrac{m}{8\mu\underline{\omega}}\underline{t}^d + \dfrac{m}{8\mu\underline{\omega}}\left(k + \dfrac{8\mu}{3}\right)\theta\dfrac{\underline{\omega x}}{m} & \text{plane stress} \\ \dfrac{m}{8\mu\underline{\omega}}\underline{t}^d + \dfrac{m}{8\mu\underline{\omega}}\dfrac{10\mu}{3}\theta\dfrac{\underline{\omega x}}{m} & \text{plane strain} \end{cases} \quad (36)$$

From (Silling et al., 2007), we know that $\underline{e}^d$ is independent of $\theta$ and $\underline{e}^i$ is independent of $\underline{t}^d$. We use this idea to decompose the extension state above (Eq. (36)), which is also followed by (Madenci and Oterkus, 2016). From Eqs. (28) and (36), the relation for the deviatoric extension state is:

$$\underline{e}^d = \left(\dfrac{m}{8\mu\underline{\omega}}\right)\underline{t}^d = \begin{cases} \underline{e} - \left(\dfrac{1}{3} + \dfrac{k}{8\mu}\right)\theta\underline{x} & \text{plane stress} \\ \underline{e} - \dfrac{5}{12}\theta\underline{x} & \text{plane strain} \end{cases} \quad (37)$$

As a result, the isotropic part of the extension state is:

$$\underline{e}^{iso} = \begin{cases} \left(\dfrac{1}{3} + \dfrac{k}{8\mu}\right)\theta\underline{x} & \text{plane stress} \\ \dfrac{5}{12}\theta\underline{x} & \text{plane strain} \end{cases} \quad (38)$$

While the formulas for the full 2D force and extension states in (Le and Bobaru, 2018) are the same as the ones we used here, the decomposition we obtained for the extension state is different from one shown in (Le and Bobaru, 2018), where the deviatoric force state is not a scalar multiple of the deviatoric extension state. This is inconsistent with the classical theory where $\sigma^d = 2G * \epsilon^d$ (Boresi et al., 1985).

The deviatoric force state for both plane stress and strain given by Eq. (28) can be rewritten in terms of $\underline{e}^d$ as:

$$\underline{t}^d = \dfrac{8\mu}{m}\underline{\omega}\underline{e}^d \quad (39)$$

where $\underline{e}^d$ is defined in Eq. (37).

## 4. Ordinary state-based peridynamics for elastoplastic modeling in 2D

In this section, a 2D elastic-perfectly plastic material model in PD is presented. Similar to the models suggested in other works (e.g. (Madenci and Oterkus, 2016; Mitchell, 2011; Silling et al., 2007)), we assume that only shear deformation is responsible for plastic deformation. The PD model introduced below is consistent with the classical rate independent J2 plasticity. The key ingredients to build this model is a constitutive equation, the yield function, a flow rule, a consistency condition, and a numerical method (return mapping algorithm) in the PD framework.

### 4.1. Peridynamic constitutive model for elastic-perfectly plastic material

Here we use the non-local elastic-plastic constitutive model proposed in (Silling et al., 2007) in 3D to derive the new 2D model. Having the consistent decomposition of extension state and force state (obtained in the previous section) is significant in defining such a PD constitutive model. According to J2 plasticity only the deviatoric part is responsible for the plastic deformation hence, the deviatoric part of extension state ($\underline{e}^d$) is decomposed into the elastic and plastic parts ($\underline{e}^{de}$ and $\underline{e}^{dp}$ respectively):

$$\underline{e} = \underline{e}^d + \underline{e}^{iso} \tag{40}$$
$$\underline{e}^d = \underline{e}^{de} + \underline{e}^{dp}$$

We express the force state for 2D plane stress and strain in isotropic elastic materials given in Eq. (13) in terms of $\underline{e}^d$ and $\theta$ as:

$$\underline{t} = 2k\theta \frac{\omega \underline{x}}{m} + \frac{8\mu}{m} \omega \underline{e}^d \tag{41}$$

Using an additive decomposition of the extension state, the PD elastoplastic constitutive model is:

$$\underline{t} = 2k\theta \frac{\omega \underline{x}}{m} + \frac{8\mu}{m} \omega (\underline{e}^d - \underline{e}^{dp}) \tag{42}$$

where $\theta$ is defined as:

$$\theta = \begin{cases} \frac{2(2\nu-1)}{(\nu-1)m} \underline{\omega x} \bullet (\underline{e} - \underline{e}^{dp}) & \text{plane stress} \\ \frac{2}{m} \underline{\omega x} \bullet (\underline{e} - \underline{e}^{dp}) & \text{plane strain} \end{cases} . \tag{43}$$

### 4.2. Yield criteria for the 2D PD model

Analogous to classical plasticity, a yield function is required to separate the elastic and plastic regions. Two approaches have been used in the literature to construct a PD yield function: (1) Based on a criterion for the deviatoric force state in 3D (Mitchell, 2011); and (2) Based on a criterion for deviatoric PD strain energy density for both 2D and 3D (Madenci and Oterkus, 2016). In both approaches, the yield function is consistent with the rate-independent von-Mises yield criterion. We discuss both approaches to construct corresponding 2D peridynamic yield functions.

### 4.2.1. A yield function based on the deviatoric strain energy density

In classical J2 plasticity, the von-Mises yield criterion is known as the maximum deviatoric strain energy criterion. The reason is the following relation between $J_2$ invariant and the distortion part of strain energy ($W^d$) (Borja, 2013):

$$\left. \begin{array}{l} W^d = \dfrac{J_2}{2\mu} \\ J_2 = \dfrac{\sigma_{vm}^2}{3} \end{array} \right\} \rightarrow W^d = \dfrac{\sigma_{vm}^2}{6\mu} \rightarrow \sigma_{vm}^2 = 6\mu W^d \qquad (44)$$

where $\sigma_{vm}$ denotes the Von-Mises stress. The yield function based on deviatoric strain energy for a perfectly plastic material is:

$$F(W^d) = \psi(W^d) - \psi_0 = W^d - \dfrac{\sigma_y^2}{6\mu} \qquad (45)$$

The peridynamic strain energy density is defined in accordance with the classical one (Le et al., 2014). The 2D strain energy density is given in (Le et al., 2014) in terms of $\underline{e}$ and $\theta$ is:

$$W(\theta, \underline{e}) = k'' \dfrac{\theta^2}{2} + \dfrac{\alpha}{2} \underline{\omega e} \bullet \underline{e}, \quad \alpha = \dfrac{8\mu}{m} \qquad (46)$$

$$k'' = \begin{cases} \dfrac{3k(1-3\nu)(1-\nu)}{2(2\nu-1)(1+\nu)} & \text{plane stress} \\ \dfrac{3k(4\nu-1)}{2(\nu+1)} & \text{plane strain} \end{cases}$$

We need to determine the deviatoric part of the total strain energy density in order to use it in the yield function. According to the classical theory, the isotropic part of strain energy density is (Abeyaratne, 2012):

$$W^{iso} = \dfrac{1}{2} k \theta^2, \qquad (47)$$

hence, the remaining part of strain energy is the deviatoric part:

$$W^d = W - W^{iso} = (k'' - k) \dfrac{\theta^2}{2} + \dfrac{\alpha}{2} \underline{\omega e} \bullet \underline{e} \qquad (48)$$

Using obtained $W^d$ in Eq. (45), gives the yield criterion in terms of distortional strain energy density for the 2D plane stress/strain conditions. The yield function in Eq. (45) is different from the corresponding one in (Madenci and Oterkus, 2016) different underlying assumptions: the formulation in (Madenci and Oterkus, 2016) is limited to no/infinitesimal rotations, whereas our model allows arbitrary rotations (see a discussion for the elastic part in (Le and Bobaru, 2018)).

### 4.2.2. A yield function based on deviatoric force state

We can alternatively construct a 2D yield function consistent with J2 plasticity based on a criterion on deviatoric force state similar to (Mitchell, 2011). In (Mitchell, 2011), the 3D yield function for $\omega = 1$ is:

$$f(\underline{t}^d) = \psi(\underline{t}^d) - \psi_0 = \frac{\|\underline{t}^d\|^2}{2} - \psi_0 \tag{49}$$

$\psi_0$ is the positive constant scalar and determines the yielding threshold. If one directly adopts Eq. (49) for 2D and follows the calibration procedure in pure shear loading mode as described in (Mitchell, 2011) (see Appendix B), $\psi_0$ is obtained to be:

$$\psi_0 = \frac{8\sigma_y^2}{3\pi l_z \delta^4} \quad \text{for } \omega = 1 \tag{50}$$

where $l_z$ is the finite thickness of the plate. Note that $\frac{\|\underline{t}^d\|^2}{2}$ in Eq. (49) is equivalent to the second invariant of the deviatoric stress tensor in classical theory ($J_2 = \frac{1}{2}\sigma^d : \sigma^d$). However, our 2D $\underline{t}^d$ ignores the effect of out of plane component of the deviatoric stress tensor ($\sigma_{33}^d$) present in the $J_2$ relation for plane stress/strain conditions. In other words, $\frac{\|\underline{t}_{2D}^d\|^2}{2}$ is equivalent to $\frac{1}{2}\sigma_{2D}^d : \sigma_{2D}^d$, where $\sigma_{2D}^d$ is the classical in-plane deviatoric stress tensor. Therefore, we need to modify the yield function for our 2D case. We modify the proposed yield function by adding the "missing" out-of-plane contribution:

$$f(\underline{t}^d) = \frac{\|\underline{t}^d\|^2}{2} + 4\frac{(\underline{t}^d \bullet \underline{x})^2}{\pi \delta^4 l_z} - \psi_0 \tag{51}$$

where $\psi_0$ is given in Eq. (50). The justification for the additional out-of-plane contribution $4\frac{(\underline{t}^d \bullet \underline{x})^2}{\pi \delta^4 l_z}$ is provided in Appendix C.

### 4.3. Equivalent Von-Mises Stress in 2D peridynamics

In this section we find the PD von-Mises equivalent stress relation for the 2D case. A similar approach has been used in (Madenci and Oterkus, 2016; Mitchell, 2011) for the 3D case.

Using the yield functions introduced in section 4.2.2, which are equivalent to J2 plasticity ($f = \sigma_{vm} - \sigma_y$), we have the following relations for the von-Mises equivalent stress:

$$\sigma_{vm}^{PD} \quad (52)$$

$$= \begin{cases} \sqrt{\dfrac{3\pi h \delta^4}{8}\left(\dfrac{\|\underline{t}^d\|^2}{2} + 4\dfrac{(\underline{t}^d \bullet \underline{x})^2}{\pi \delta^4 h}\right)} & \text{with the yield function based on } \underline{t}^d(\text{Eq. (51)(51)}) \\ \sqrt{6\mu W^d} & \text{with the yield function based on } W^d(\text{Eq. (45)(44)}) \end{cases}$$

Based on the classical definition we also have:

$$\sigma_{vm} = \sqrt{\frac{1}{2}[(\sigma_{11} - \sigma_{22})^2 + (\sigma_{22} - \sigma_{33})^2 + (\sigma_{33} - \sigma_{11})^2 + 6\sigma_{12}^2]} \quad (53)$$

where $\sigma_{ij}$'s are the stress components of the PD collapsed stress tensor (see Eq. (14)). Recall that 4.2.2 $\sigma_{31} = \sigma_{32} = 0$ in 2D, and $\sigma_{33}$ is zero for plane stress, but a function of $\sigma_{11}$ and $\sigma_{22}$ for plane strain conditions.

Comparing the $\sigma_{vm}^{PD}$ values (Eq. 52, which are obtained from force states directly) with the ones obtained based on the classical definition given in Eq. (53) (obtained from conversion of states to tensors and then using the classical definition of the von-Mises equivalent stress) is one way to verify the consistency of the proposed yield functions with classical J2 plasticity.

### 4.4. Plastic flow rule and consistency condition

In classical plasticity, the plastic flow rule is used to update the plastic part of strain and accordingly the stress when the yield condition is reached. The flow rule determines the plastic flow direction. A PD plastic flow rule suggested in (Mitchell, 2011) describes the rate of plastic extension as follows:

$$\dot{e}^{dp} = \lambda \nabla^d \psi \quad (54)$$

where $\lambda$ is the consistency parameter (a scalar), and $\nabla^d \psi$ is Fréchet derivative of $\psi$ with respect to $\underline{t}^d$.

In classical plasticity, loading /unloading and consistency are governed by the Kuhn-Tucker conditions. According to (Mitchell, 2011), *Kuhn- Tucker conditions* for PD can be expressed as:

$$\lambda \geq 0, \ f(\underline{t}^d) \leq 0, \lambda f(\underline{t}^d) = 0 \quad (55)$$

meaning:
  a) $f(\underline{t}^d) < 0 \rightarrow \lambda = 0$ elastic
  b) $f(\underline{t}^d) = 0 \rightarrow \lambda \geq 0$ yield

In the case that the yield condition (b) is reached, the following *consistency condition* is considered for finding $\lambda$:

$$\lambda \dot{f}(\underline{t}^d) = 0 \quad (56)$$

meaning:

a) $\lambda > 0 \rightarrow \dot{f}(\underline{t}^d) = 0$    plastic loading

b) $\dot{f}(\underline{t}^d) < 0 \rightarrow \lambda = 0$    unloading (elastic)

In the case of using deviatoric strain energy density for yield function, $f(\underline{t}^d)$ should be replaced with $F(W^d)$ in Eqs. (55) and (56).

In the numerical solution of a plasticity problem, these conditions together with the flow rule and a return mapping algorithm are used to determine the plastic extension and force states.

### 4.5. The return mapping algorithm

*Return mapping* (RM) algorithms are a class of numerical algorithms that are used to determine the plastic part of deformation and update other quantities in an incremental loading process. Here we follow the idea for the peridynamic RM algorithm presented in (Mitchell, 2011) (for the 3D case) to develop a 2D RM algorithm using the 2D force and extension states obtained above. Our algorithm is consistent with the classical flow rule, which is different from the algorithm proposed in (Madenci and Oterkus, 2016). This approach results in a nonlinear system of equations, with the number of unknowns/equations being equal to the number of family nodes. We reduce this system to a single nonlinear equation (with one unknown) by performing linear algebra operations on the discrete version of the states. This allows the algorithm to remain efficient while being consistent with the classical flow rule.

Given $\underline{e}_n^d$, $\underline{e}_n^{dp}$ from the *n*-th increment and a new deformation increment $\Delta \underline{e}$, the new plastic deformation ($\underline{e}_{n+1}^{dp}$) and force state $\underline{t}_{n+1}$ need to be determined. In RM algorithms, we first compute a "trial" force state ($\underline{t}_{trial}^d$) assuming that the total new extension increment is elastic. Then, if the yield criterion is violated, the flow rule and consistency condition are used to correct the trial value and find the amount of plastic extension.

$$\underline{e}_{trial} = \underline{e}_{n+1} - \underline{e}_n^{dp} \tag{57}$$

where $\underline{e}_{trial}$ is the trial elastic extension, assuming $\Delta \underline{e}$ is all elastic. Then from Eq. (28):

$$\underline{t}_{trial}^d = \begin{cases} \left(-k - \dfrac{8\mu}{3}\right) \theta(\underline{e}_{trial}) \dfrac{\omega x}{m} + \dfrac{8\mu}{m} \underline{\omega}\, \underline{e}_{trial} & \text{plane stress} \\ -\dfrac{10\mu}{3} \theta(\underline{e}_{trial}) \dfrac{\omega x}{m} + \dfrac{8\mu}{m} \underline{\omega}\, \underline{e}_{trial} & \text{plane strain} \end{cases} \tag{58}$$

where similar to Eq. (12):

$$\theta(\underline{e}_{\text{trial}}) = \begin{cases} \dfrac{2(2\nu - 1)}{(\nu - 1)m} \underline{\omega x} \bullet \underline{e}_{\text{trial}} & \text{plane stress} \\ \dfrac{2}{m} \underline{\omega x} \bullet \underline{e}_{\text{trial}} & \text{plane strain} \end{cases} \qquad (59)$$

Then, the yield condition is checked to determine whether the trial values require correction:

- $f(\underline{t}_{\text{trial}}^d) \leq 0 \rightarrow \lambda = 0 \rightarrow$ elastic step $\rightarrow \begin{cases} \underline{t}_{n+1}^d = \underline{t}_{\text{trial}}^d \\ \underline{e}_{n+1}^{\text{dp}} = \underline{e}_n^{\text{dp}} \end{cases}$ (60)
- $f(\underline{t}_{\text{trial}}^d) > 0 \rightarrow \lambda > 0 \rightarrow$ incrementally plastic, use consistency

According to Eq. (60) if the yield condition is not reached, the trial states are valid quantities of that increment. Otherwise, plasticity is involved and the consistency condition is used to find the $\lambda$ value to update the plastic extension state, and to correct the trial force state accordingly.

Note that we proceed with the force-state yield function given by Eq. (51). However, the following approach is general, and the proposed steps are applicable to the strain energy density yield function as well.

If the second condition in Eq. (60) holds, based on the consistency condition in Eq. (56) we get:

$$f(\underline{t}_{n+1}^d) = 0 \rightarrow \frac{\|\underline{t}_{n+1}^d\|^2}{2} + 4 \frac{(\underline{t}_{n+1}^d \bullet \underline{x})^2}{\pi \delta^4 l_z} - \psi_0 = 0; \qquad \text{for } \omega = 1 \qquad (61)$$

The 2D scalar deviatoric force state given in Eq. (28), can now be written in terms of $\underline{t}_{\text{trial}}^d$ and the consistency parameter (see Appendix D for derivation) as follows:

$$\underline{t}_{n+1}^d \qquad (62)$$
$$= \begin{cases} \underline{t}_{\text{trial}}^d - \dfrac{8\mu}{m} \underline{\omega}(\Delta\lambda \nabla^d \psi) + \dfrac{2(2\nu - 1)}{\nu - 1}\left(k + \dfrac{8\mu}{3}\right)\dfrac{\underline{\omega x}}{m}\left(\dfrac{\underline{\omega x}}{m} \bullet \Delta\lambda \nabla^d \psi\right) & \text{plane stress} \\ \underline{t}_{\text{trial}}^d - \dfrac{8\mu}{m} \underline{\omega}(\Delta\lambda \nabla^d \psi) + \dfrac{20\mu}{3}\dfrac{\underline{\omega x}}{m}\left(\dfrac{\underline{\omega x}}{m} \bullet \Delta\lambda \nabla^d \psi\right) & \text{plane strain} \end{cases}$$

The discrete consistency parameter is $\Delta\lambda$ (Simo and Hughes, 2006) and $\nabla^d \psi = \underline{t}^d$ according to Eq. (61). Here we refer to Eq. (62) as the *trial-corrector* equation. The detailed derivation of Eq. (62) and $\nabla^d \psi$ is provided in Appendix D. The consistency condition in Eq. (61) and the trial-corrector in Eq.(62), lead to a system of equations with the unknowns being $\Delta\lambda$ and $\underline{t}_{n+1}^d$. At the continuous level, each point has an infinite dimensional $\underline{t}_{n+1}^d$ (due to the infinite number of points in the family of a node). After discretization, $\underline{t}_{n+1}^d$ becomes $m$-dimensional with $m$ being the number of family nodes. Then, Eq.(62) leads to a system with $m+1$ equations and $m+1$ unknowns (the consistency parameter and the $m$ magnitudes of the deviatoric force state). By comparison, in the classical theory the corresponding system of equations has 4 unknowns the consistency parameter and the 3 components of the deviatoric stress tensor (Simo and Hughes, 2006).

Note that the trial-corrector equation is implicit in terms of $\underline{t}_{n+1}^d$. A similar system of equations also emerges for the 3D formulation, see (Mitchell, 2011). In the 3D case (Mitchell, 2011), however, the RM system is linear and can be easily solved analytically (Mitchell, 2011). For the 2D cases observed here and also in (Madenci and Oterkus, 2016), all equations in the system are nonlinear. For the 2D classical formulation, the RM system is also nonlinear. However, since classical formulations are based on tensors, linear algebra is used to reduce the system into one nonlinear equation with a single unknown: $\Delta\lambda$ (Simo and Hughes, 2006). For the 2D peridynamic formulation, such reduction of the system into one equation is not as straightforward since state operations with integral operators are involved.

In (Madenci and Oterkus, 2016) authors used an assumption to reduce the system of equation to one nonlinear scalar equation. Instead of using the true flow rule, they used an equation, which is similar to using $\Delta\underline{e}_{n+1}^{dp} = \Delta\lambda \underline{t}_{trial}^d$ instead of $\Delta\underline{e}_{n+1}^{dp} = \Delta\lambda \underline{t}_{n+1}^d$ in our model. However, this approach can potentially compromise accuracy if $\underline{t}_{n+1}^d$ and $\underline{t}_{trial}^d$ differ from each other significantly. Here, we reduce the system to a single nonlinear equation while utilizing the true flow rule: $\Delta\underline{e}_{n+1}^{dp} = \Delta\lambda \underline{t}_{n+1}^d$.

We first show how to derive the explicit trial-corrector relationship. For simplicity, we consider $\omega = 1$. However, the approach is general and can be used for any influence function. To simplify notation, we define the following:

$$A = \frac{4}{\pi \delta^4 h} \tag{63}$$

$$B = -\frac{8\mu}{m}\underline{\omega}$$

$$C = \begin{cases} \dfrac{2(2\nu - 1)}{\nu - 1}\left(k + \dfrac{8\mu}{3}\right) & \text{plane stress} \\ \dfrac{20\mu}{3} & \text{plane strain} \end{cases}$$

The system of equations (consistency condition and trial-corrector function) becomes:

$$\begin{cases} \dfrac{\left\|\underline{t}_{n+1}^d\right\|^2}{2} + A(\underline{t}_{n+1}^d \bullet \underline{x})^2 - \psi_0 = 0 \\ \underline{t}_{n+1}^d = \underline{t}_{trial}^d + B\Delta\lambda \underline{t}_{n+1}^d + C\Delta\lambda \underline{x}\ (\underline{x}\bullet\underline{t}_{n+1}^d) \end{cases} \tag{64}$$

The discretized version of this implicit trial-corrector equation is:

$$\boldsymbol{t}_{n+1,i}^d = \boldsymbol{t}_{trial,i}^d + B\Delta\lambda \boldsymbol{t}_{n+1,i}^d + C\Delta\lambda \boldsymbol{x}_i\big[\boldsymbol{x}_i \cdot \big(\boldsymbol{t}_{n+1,i}^d V_i\big)\big] \tag{65}$$

If each node $i$ has $m$ nodes in its family, $\boldsymbol{t}^d_{n+1,i}$, $\boldsymbol{t}^d_{trial,i}$, and $\boldsymbol{x}_i$ are $m-$ dimensional vectors store the values of the vector states $\underline{t}^d_{n+1}$, $\underline{t}^d_{trial}$ and $\underline{x}$ at this node. $\boldsymbol{V}_i$ is also an $m$-dimensional vector storing the partial volumes (areas) for the family nodes of the $i^{th}$ node. In Eq. (65), ($\cdot$) is the vector dot product operator, while $\boldsymbol{t}^d_{n+1,i}\boldsymbol{V}_i$ is the element-wise product of the two vectors. Reorganizing Eq. (65), we obtain:

$$\boldsymbol{t}^d_{trial,i} = (1 - B\Delta\lambda)\boldsymbol{t}^d_{n+1,i} - C\Delta\lambda[\boldsymbol{x}_i \otimes (\boldsymbol{x}_i \boldsymbol{V}_i)] \cdot \boldsymbol{t}^d_{n+1,i} \tag{66}$$

Defining:

$$\boldsymbol{M}_i = (1 - B\Delta\lambda)\boldsymbol{I} - C\Delta\lambda[\boldsymbol{x}_i \otimes (\boldsymbol{x}_i \boldsymbol{V}_i)] \tag{67}$$

with $\boldsymbol{I}$ being the identity matrix, we find:

$$\boldsymbol{t}^d_{n+1,i} = \boldsymbol{M}_i^{-1} \boldsymbol{t}^d_{trial,i} \tag{68}$$

Using the Sherman-Morrison formula we obtain:

$$\boldsymbol{M}_i^{-1} = \frac{1}{1 - B\Delta\lambda}\boldsymbol{I} + \frac{C\Delta\lambda}{(1 - B\Delta\lambda)\{1 - \Delta\lambda[B + C\boldsymbol{x}_i \cdot (\boldsymbol{x}_i \boldsymbol{V}_i)]\}} \boldsymbol{x}_i \otimes (\boldsymbol{x}_i \boldsymbol{V}_i) \tag{69}$$

According to Eqs. (68) and (69):

$$\boldsymbol{t}^d_{n+1,i} = \frac{1}{1 - B\Delta\lambda}\boldsymbol{t}^d_{trial,i} + \frac{C\Delta\lambda}{(1 - B\Delta\lambda)\{1 - \Delta\lambda[B + C\boldsymbol{x}_i \cdot (\boldsymbol{x}_i \boldsymbol{V}_i)]\}} \boldsymbol{x}_i \otimes (\boldsymbol{x}_i \boldsymbol{V}_i) \boldsymbol{t}^d_{trial,i} \tag{70}$$

Having the explicit discrete version of the trial-corrector in Eq. (70) we can write the continuous version as well:

$$\underline{t}^d_{n+1} = \frac{1}{1 - B\Delta\lambda}\underline{t}^d_{trial} + \frac{C\Delta\lambda}{(1 - B\Delta\lambda)\left\{1 - \Delta\lambda\left[B + C\|\underline{x}\|^2\right]\right\}} \underline{x}\ (\underline{x} \bullet \underline{t}^d_{trial}) \tag{71}$$

Note that deriving Eq. (71) from Eq. (64)(62) is not apparent. The fact that we moved to the discretization form allowed us to use simple linear algebra operations to obtain an explicit description for $t^d_{n+1,i}$, and transform back to the continuous (in space) form to obtain $\underline{t}^d_{n+1}$ explicitly.

Substituting Eq. (71) into the yield function in Eq. (64)-first equation, results in the following equation in terms of $\Delta\lambda$ being the unknown:

$$G[P_1 + 2AP_2] + H^2[P_2P_3(1 + 2AP_3)] + 2GH[P_2 + 2AP_2P_3] - 2\psi_0 = 0 \tag{72}$$

where

$$P_1 = \|\underline{t}_{\text{trial}}^d\|^2; \quad P_2 = (\underline{x} \bullet \underline{t}_{\text{trial}}^d)^2; \quad P_3 = \|\underline{x}\|^2$$

$$G = \frac{1}{1 - B\Delta\lambda}; \quad H = \frac{C\Delta\lambda}{(1 - B\Delta\lambda)[1 - \Delta\lambda(B + CP_3)]}$$

and $A$, $B$, and $C$ given by Eq.(63).

The system of equation in Eq. (64) is reduced to Eq. (72), a nonlinear scalar equation. We solve Eq. (72) via Newton-Raphson method to find $\Delta\lambda$. Then $\underline{t}_{n+1}^d$ is updated using Eq. (71), and $\underline{e}_{n+1}^{dp}$ is updated with:

$$\underline{e}_{n+1}^{dp} = \underline{e}_n^{dp} + \Delta\lambda \nabla^d \psi = \underline{e}_n^{dp} + \Delta\lambda \underline{t}_{n+1}^d \tag{73}$$

$$\underline{e}_{n+1}^e = \underline{e}_{n+1} - \underline{e}_{n+1}^{dp}$$

According to Eq. (43) :

$$\theta_{n+1} = \begin{cases} \frac{2(2\nu-1)}{(\nu-1)m} \underline{\omega x} \bullet \underline{e}_{n+1}^e & \text{plane stress} \\ \frac{2}{m} \underline{\omega x} \bullet \underline{e}_{n+1}^e & \text{plane strain} \end{cases} \tag{74}$$

Following the procedure above, a similar return mapping algorithm can be constructed for the case when the yield function is based on strain energy density. For convenience, a step-by-step outline of the overall computational scheme is given in Table 1:

Table 1.   Return-Mapping Algorithm for rate-independent, perfect plasticity PD model

1. Given data: $\{\underline{e}_{n+1}, \underline{e}_n^d, \underline{e}_n^{dp}\}$
2. Compute elastic trial deviatoric force state and test for plastic loading

$$\underline{e}_{\text{trial}} = \underline{e}_{n+1} - \underline{e}_n^{dp} \rightarrow \underline{t}_{\text{trial}}^d \rightarrow f(\underline{t}_{\text{trial}}^d)$$

   If $f(\underline{t}_{\text{trial}}^d) \leq 0$ then
   (2a)  Elastic step: $\Delta\lambda = 0$, set $(\ )_{n+1} = (\ )_{\text{trial}}$ and exit
   Else
   (2b)  Plastic step: $\Delta\lambda > 0$, proceed to step 3
   End if
3. Return mapping
   Find $\Delta\lambda$ using the Newton-Raphson method

Update: $\underline{t}_{n+1}^d$ , $\underline{e}_{n+1}$

## 5. Numerical discretization and numerical solution

Various numerical schemes can be used for PD models (Dipasquale et al., 2014; Emmrich and Weckner, 2007; Hu et al., 2010; Jafarzadeh et al., 2020; Silling and Askari, 2005). In this study, the meshfree discretization scheme (Silling and Askari, 2005) is used due to certain advantages it offers when applied to damage and fracture problems, which we will consider in a follow-up paper. For the spatial integration of Eq.(8), we use the one-point Gaussian quadrature rule. For convenience, the domain is discretized with a uniform square grid with $\Delta x$ grid spacing (see Figure 4).

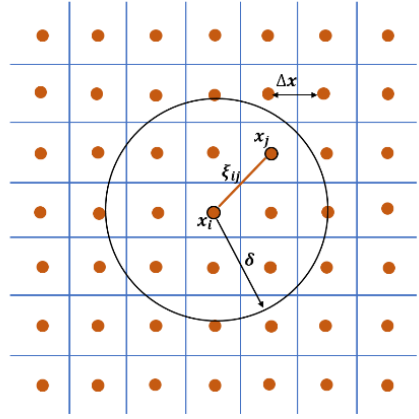

Figure 4: The uniform 2D discretization. In practice, a ratio $\delta/\Delta x$ of 4 or more is used to insure mesh-independency (see (Chen et al., 2016; Dipasquale et al., 2014)).

The spatially discretized version of Eq. (8) is then:

$$\rho(x_i)\ddot{u}(x_i,t) = \int_{\mathcal{H}_x} (\underline{T}[x_i,t]\langle x'-x_i\rangle - \underline{T}[x',t]\langle x_i-x'\rangle)dA_{x'} \qquad (75)$$
$$+ b(x_i,t)$$
$$\approx \sum_j \left(\underline{T}[x_i,t]\langle x_j-x_i\rangle - \underline{T}[x_j,t]\langle x_i-x_j\rangle\right)A_{ij} + b(x_i,t)$$

where $V_{ij}$ is the portion of volume of the node $x_j$ covered by horizon region of the node $x_i$. To improve the accuracy of the numerical integration scheme in PD and reduce the discretization effect, we use the IAP-3 algorithm proposed in (Hu et al., 2010) (which is called PA-HBB in (Seleson, 2014)). The algorithm approximates $V_{ij}$ for the $x_j$ that are not fully covered by the horizon of $x_i$.

In a quasi-static analysis, the acceleration term in the left-hand side of the above equation is set to zero. We used the nonlinear conjugate gradient method (Shewchuk, 1994) to find the equilibrium solution for quasi-static analysis. Dynamic relaxation (Kilic and Madenci, 2009) and direct methods (Sarego et al., 2016) are other methods that can be used in a PD quasi-static analysis. As was explained in the previous section, we use a return mapping algorithm to update the plastic deformation. The algorithm for the quasi-static elastoplastic analysis is as follows:

A) Read input data (material properties, initial and boundary conditions)
B) Update displacement boundary conditions at the corresponding load step.
C) At load step $n$, decompose the extension state into deviatoric and dilatation parts based on Eqs. (37) and (38).
D) Find the equilibrium solution (displacement field) using the nonlinear conjugate gradient (CG) method. Note that in each iteration of CG, the RM algorithm (Table 1) is used to update the plastic values of extension and force states based on the new displacements that CG suggested. The norm of the nodal residual force is used as a stopping criterion.

   iterate until the norm of nodal residual forces becomes smaller than a certain tolerance (we choose $10^{-6}$ in the examples below). Then go to next load step (B).

E) End

## 6. Results and discussion

In this section, we solve two plane stress peridynamic examples using a PD code implemented in MATLAB. The simulation results with the proposed 2D-PD elastoplastic model are compared to the results based on the local J2 plasticity obtained with Abaqus.

### 6.1. Dog-bone specimen under tension

For the first example, we use a dog-bone specimen. The thin plate is 50 mm by 100 mm and the curvatures of 10mm. See Figure 5 for geometry and boundary conditions. The material parameters in this example, Young's modulus, Poisson's ratio, yield stress, and density are respectively set to be: $E$ = 91 GPa, $\nu$ = 0.33, $\sigma_y = 300$ MPa, and $\rho = 1049$ kg/m$^3$, representative of some aluminum alloys.

The bottom of the plate is fixed in the y-direction and a uniaxial total displacement loading $u_y = 0.35$mm is imposed in the PD simulation through 25 equally spaced loading steps on the top boundary. Different schemes for imposing nonlocal boundary conditions (BCs) are shown in, e.g., (Kilic and Madenci, 2009), (Madenci and Oterkus, 2014), (Aksoylu et al., 2018). When comparing results with classical solutions, one tries to mimic the imposition of the corresponding local BCs and decrease the peridynamic surface effect (Le and Bobaru, 2017). Here, we use the fictitious nodes method (Oterkus et al., 2014) for applying the Dirichlet boundary conditions (BCs). As seen in Figure 5, the displacements are applied over a layer of fictitious nodes on the top and bottom boundary layers of the sample with its width equal to the horizon size ($\delta$). Applying BC with this method is equivalent to imposing local BC on the boundaries (dash-line in Figure 5) which are used in our local FEM analysis. In the PD simulations, we used a horizon size of $\delta = 2.5$ mm and the ratio $\frac{\delta}{\Delta x} = 5$ (a convergence study in terms of the horizon size is shown below). The total number of nodes for this horizon size is then 20,168.

In our finite element simulation with Abaqus, we used a finer mesh (66,670 linear quadrilateral elements type CPS4R), which has a relative difference of less than 0.01% for the magnitude (in norm-2) of the displacement vector at one point along the horizontal symmetry line, near the edge of the sample, when computed with two different meshes, one having twice the number of elements. Both the PD and Abaqus analyses are quasi-static. Note that using the same 25 equally spaced load increments in the Abaqus model as in the PD simulation leads to an error message in Abaqus, complaining about the increment size being too large. The same message is produced when using ten times the number of increments. We finally use 500 load increments in Abaqus to obtain the results shown below. Part of the differences between the PD and Abaqus results are caused by the difference in the loading steps used to obtain those results.

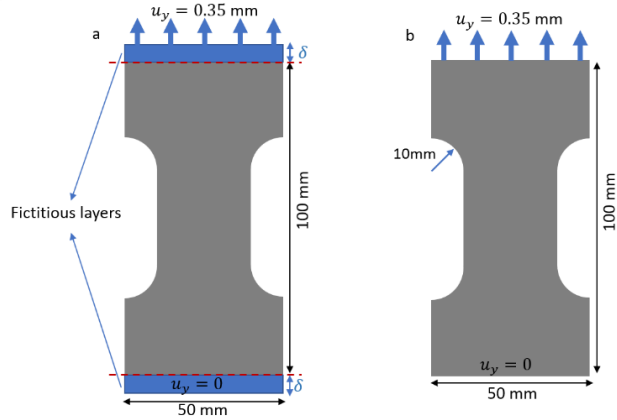

Figure 5: Sample geometry and boundary conditions in PD (a), FEM /Abaqus (b)

In Figure 6, we compare the displacements at the end of the applied loading obtained by the 2D PD model with the force state yield function (see Section 4.2.2), and by the Abaqus elastic-perfectly plastic analysis. The PD displacements are in good agreement with the local J2 plasticity results. Some minor differences are seen near the side boundaries, likely caused by the peridynamic surface effect (see (Le and Bobaru, 2017)). Different schemes for reducing or eliminating the peridynamic surface effect (so called "surface correction algorithms") have appeared in (Emmrich and Weckner, 2007; Le and Bobaru, 2017; Mitchell et al., 2015; Oterkus et al., 2014). Here, we used the fictitious nodes method (Oterkus et al., 2014) to apply the boundary conditions on the top and bottom sides of the dog-bone specimen. We do not do anything special for the load-free boundaries, where the surface effect is visible at a closer examination.

**Error! Reference source not found.** shows the von-Mises stress obtained by Abaqus, in comparison with the Von-Mises equivalent stress from the PD simulation with both yield functions introduced in section 0.

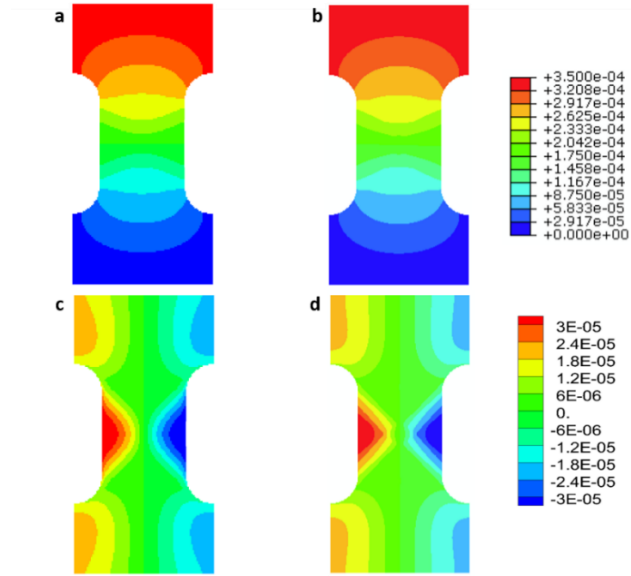

Figure 6. Vertical displacements obtained with PD (a), and FEM/Abaqus (b). Horizontal displacements from PD (c), and Abaqus (d).

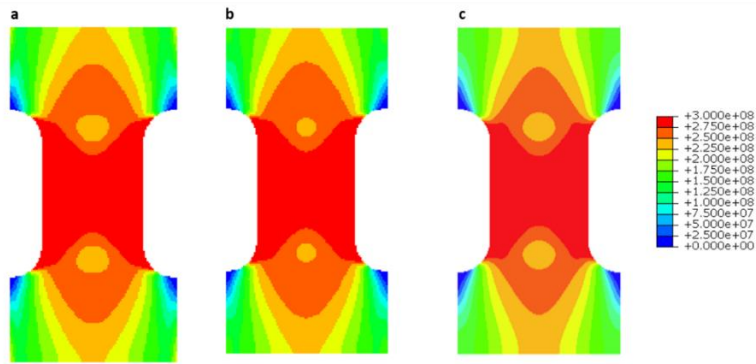

Figure 7: The von-Mises equivalent stress from the PD simulation with the yield function based on the deviatoric force state (a), and on the deviatoric strain energy density (b), and from the Abaqus solution (c).

We observe that PD results are consistent with those of the local model. Again, minor differences are due to the surface effect on the load-free boundaries in the PD models, and the difference in the size of the loading step used: Abaqus results used a loading step 20 times finer than the one we used in the PD model. We also note some small differences between the two yield functions used in the PD formulation. The von-Mises equivalent stress in **Error! Reference source not found.** is calculated via Eq. (52) in section4.3. as it is mentioned before to verify the consistency of the proposed yield functions with the local J2 plasticity and evaluate their performance, we calculate the von-Mises equivalent stress with Eq. (53) as well. This verification and further discussion are provided in Appendix E.

We also perform a $\delta$- convergence study for our model with the force-state yield function. We use three different horizon sizes, $\delta = 2.5\text{mm}, \delta = 1.25 \text{ mm}$ and $\delta = 0.625\text{mm}$ with a fixed $m = 5$. Figure 8: The PD von-Mises stress distribution with m=5 and the horizon sizes of: 2.5mm (a), (b) 1.25mm (b) 0.625 mm (c), and Abaqus solution (d). shows the von-Mises equivalent stress. As $\delta$ decreases, the PD results become smoother and in a better agreement with the local model.

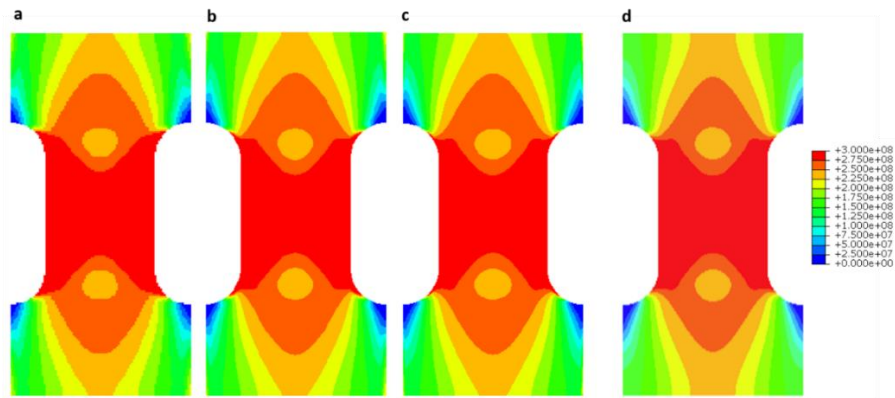

Figure 8: The PD von-Mises stress distribution with m=5 and the horizon sizes of: 2.5mm (a), (b) 1.25mm (b) 0.625 mm (c), and Abaqus solution (d).

**Error! Reference source not found.** shows specifically the plastic region where $\sigma_{vm} = 300 \text{ MPa}$ obtained by PD using 3 different horizon sizes (2.5mm, 1.25mm and 0.625mm), and by FEM (Abaqus). The match between the plastic regions from Abaqus and PD (using the smallest horizon size) is very good, in spite of the large difference in loading step between the two solution methods. The main difference between the results is the narrow plastic region bands near the side surfaces of the sample seen in the PD results and not in the Abaqus ones. These bands are caused by the PD surface effect. This effect can be eliminated by using the fictitious nodes method for arbitrary geometry algorithms recently developed in (Zhao et al., 2020). The evolution of the plastic region obtained with the PD model (using the large and small horizon sizes) and from the Abaqus simulation, can be seen in the supplementary materials in Video 1. There, the Abaqus results show a "smoother" evolution because of the 20 times finer loading step used compared with the loading step we used for the PD results. Nevertheless, not only are the final plastic regions similar between the PD and Abaqus results, but the same patterns are produced by the two different methods before reaching that final state: plastic deformations initiate at four points at the samples' edges near the part of the sample where it starts to widen.

These regions grow, in a 45° slanted direction towards the center of the sample and merge with a fifth "island" region that initiates, later, in the center of the plate in a horizontal shape that "butterflies" towards the four other plastic regions before merger.

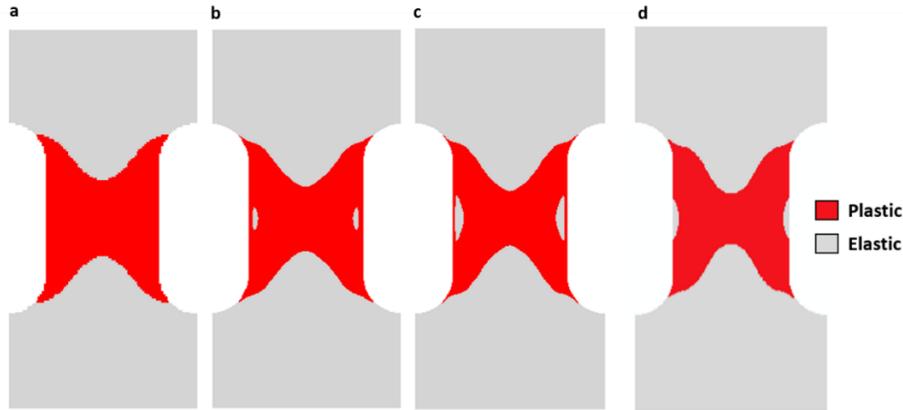

Figure 9: The extent of the plastic region computed by PD with the yield function based on deviatoric force state and horizon sizes of: 2.5 mm (a), 1.25 mm (b), 0.625 mm (c), and by Abaqus (d).

### 6.2. Cantilever beam: an example with large rotations

To show that our PD elastoplastic model works for deformations with large rotations, we simulate the bending of a cantilever beam under a vertical point load at one end and zero horizontal displacement at the other end. The cantilever beam geometry and boundary conditions are shown in Figure 10. Material properties are the same as in the previous example.

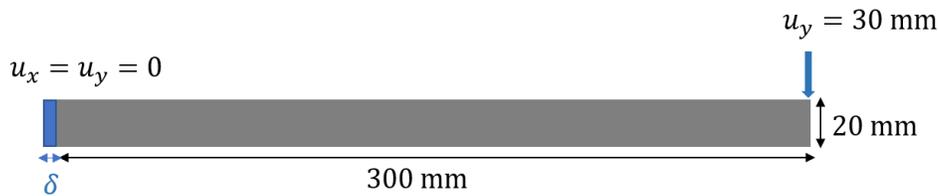

Figure 10: The 2D cantilever beam geometry and boundary conditions.

With a sufficiently large loading applied at the right end, the cantilever will experience deformations with large rotations. Using a horizon size of 2.5 mm and horizon factor $m=5$, the resulting total number of PD nodes is 25,461. We apply $u_x, u_y = 0$ on the left side of the bar over a layer of thickness equal to the horizon size and $u_y = 30$ mm is applied to the end of the bar over 30 equal increments of 1 mm.

In the corresponding FEM model, we use a finer mesh to make sure the results are converged ones (96,000, 80 by 1200 linear quadrilateral elements type CPS4R, based on a square mesh). Abaqus uses 50 increments, with steps automatically and internally determined by the nonlinear solver in the software. Figure 11 presents the FEM-based von-Mises stress and the von-Mises equivalent stress in the bar obtained via our PD model using the yield criterion based on the deviatoric force state. Figure 12 shows the extent of the plastic region obtained by PD and Abaqus. In the supplementary materials, we show the evolution of the plastic region obtained from the PD model and from Abaqus in Videos 2 and 3, respectively. Horizontal displacements obtained with both PD and FEM (Abaqus) are shown in Figure 13. Obviously, in Abaqus one needs to set the nonlinear geometry option Nlgeom (Abaqus, 2014) ON, in order to obtain correct results. In Figure 13 we also show the horizontal displacements from Abaqus with the Nlgeom setting turned OFF. Recall that in Section 3.2.1 we explained why in PD there is no need for doing anything special for correctly solving this type of large rotations problems.

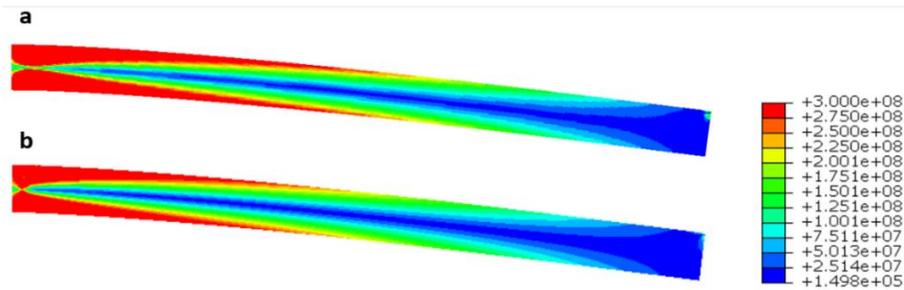

Figure 11: von-Mises equivalent stress distribution from the PD model (a), and from Abaqus (b).

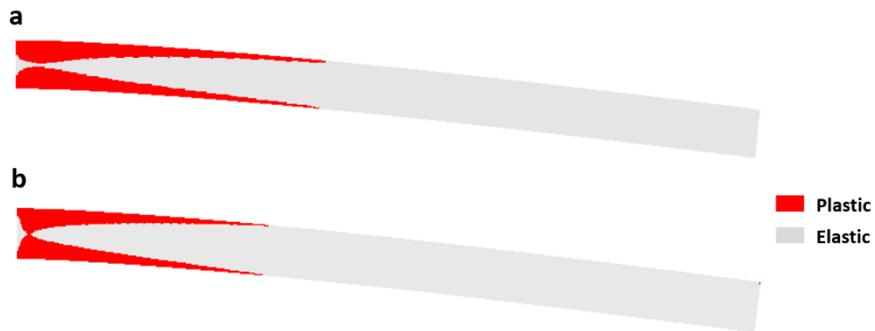

Figure 12: The extent of the plastic region for the cantilever beam obtained with PD (a), and with Abaqus (b).

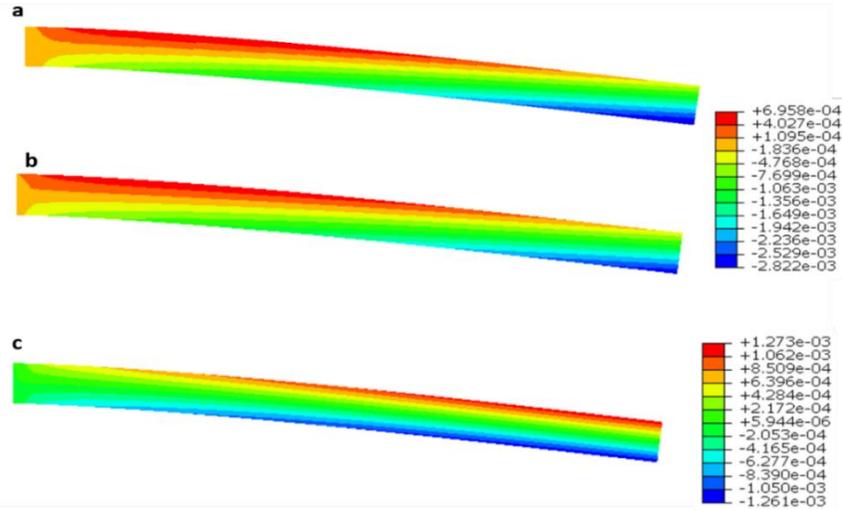

Figure 13: Horizontal displacement along the bar from PD (a), Abaqus with Nlgeom ON (b), and Abaqus with Nlgeom OFF (c).

We conclude that our elastoplastic PD model is capable of correctly capturing elastoplastic deformations with small strains and large rotations. Extending the current formulation to include large strains is left for future research and may be achieved by employing alternative strain measures for bonds (Foster and Xu, 2018; Nguyen and Oterkus, 2020).

## 7. Conclusions

In this study, a new ordinary state-based peridynamic (OSB-PD) elastoplastic model for 2D plane stress/strain conditions was introduced. The new model is capable of handling small strains and large rotations. Earlier OSB-PD 2D models for elastoplasticity were only valid for small rotations or had other inconsistencies. We showed that a new decomposition of the 2D peridynamic force and extension states was consistent with the classical isotropic-deviatoric decompositions of stress/strain tensors. The decomposition plays a key role in the model, and we provided a strategy for testing the consistency of an OSB-PD formulation for elastic-plasticity in two dimensions. Two different rate-independent yield functions consistent with J2-plasticity were introduced: one utilizing the deviatoric force-state and the other using the deviatoric PD strain energy density.

A new return-mapping algorithm is introduced for plane stress and plane strain conditions that is consistent with the classical flow rule. We reduced the resulting nonlinear system of equations to a single nonlinear equation by performing linear algebra operations on the discretized version of the PD states. This led to important gains in efficiency.

We verified the model in two computational examples and compared our solutions with those obtained from corresponding finite element results with Abaqus for elastic perfectly-plastic materials. Even when large rotations are present small-strain deformations, the newly introduced model matched the Abaqus solution (with the geometric nonlinearity option ON). We also monitored the evolution of the plastic region, which compared favorably, step-by-step, for both examples, with the Abaqus results.

In the future, the new formulation will be coupled with a damage initiation and growth model to produce a "native" (ordinary-state based) PD model for that can simulate ductile failure beyond what existing PD correspondence models are currently able to do.

## Acknowledgments

This work has been supported in part by the US National Science Foundation CMMI CDS&E Grant No. 1953346 (program manager Joanne Culbertson), by the AFOSR MURI Center for Materials Failure Prediction through Peridynamics (program managers Jaimie Tiley, David Stargel, Ali Sayir, Fariba Fahroo, and James Fillerup), and by a Nebraska Research Initiative (NRI) Systems Science grant. The computations have been performed utilizing the Holland Computing Center of the University of Nebraska, which receives support from the NRI. We would also like to acknowledge helpful discussions with Dr. John A. Mitchell from Sandia National Laboratories.

**Appendix A**

**Decomposition of the classical stress tensor**

In this part, we briefly describe the stress tensor decomposition which is used in decomposing the states in sections 3.1 and 3.2.

Under mechanical loadings, deformable bodies change their shape and volume. Two measures of such deformations are the hydrostatic and the deviatoric parts of the stress tensor.

The **hydrostatic (isotropic) stress tensor** is related to volume changes in a body. A pure hydrostatic stress tensor is (Abeyaratne, 2012):

$$\sigma^{iso} = \begin{bmatrix} \sigma_{11}^{iso} & 0 & 0 \\ 0 & \sigma_{22}^{iso} & 0 \\ 0 & 0 & \sigma_{33}^{iso} \end{bmatrix} \quad , \quad \sigma_{11}^{iso} = \sigma_{22}^{iso} = \sigma_{33}^{iso} = \frac{1}{3}\sigma_{ii} \tag{A-1}$$

$$\sigma_{ii} = \sigma_{11} + \sigma_{22} + \sigma_{33}$$

The **deviatoric stress tensor** is related to the change in body shape. The deviatoric stress tensor describes a state of pure shear, and is defined by (Abeyaratne, 2012):

$$\sigma^d = \sigma - \sigma^{iso} = \begin{bmatrix} \sigma_{11} - \sigma_{11}^{iso} & \sigma_{12} & \sigma_{13} \\ \sigma_{21} & \sigma_{22} - \sigma_{22}^{iso} & \sigma_{23} \\ \sigma_{31} & \sigma_{32} & \sigma_{33} - \sigma_{33}^{iso} \end{bmatrix} \tag{A-2}$$

One of the important properties of deviatoric stress is that its first invariant is zero.

$$J_1 = \sigma_{ii}^d = 0, \qquad \qquad J_1\text{: the first invariant of the deviatoric stress tensor} \qquad (A\text{-}3)$$
$$\sigma^d$$

## Appendix B

To derive the yield value for our 2-D yield function we used an approach the same as the one described in (Mitchell, 2011), which is for a 3D case. Similar to that approach, to find the value of $\psi_0$ we used pure shear deformation which does not contains any dilatation. Displacement field for our 2D case is:

$$u_1 = \gamma x_2 \qquad (B\text{-}1)$$
$$u_2 = 0$$

Where $\gamma$ is a shear strain, $u$ is displacement of a material point and $x$ is coordinate of the material point in reference configuration where its subscript stands for direction of the coordinate system in Cartesian coordinate. The scaler deformation state for this displacement field is written as:

$$\underline{Y}(\gamma) = \sqrt{x_2^2 + (x_1 + \gamma x_2)^2} \qquad (B\text{-}2)$$

After linearizing the Taylor expansion of the scaler deformation state, the scaler extension state is:

$$\underline{e} = |\underline{Y}| - |\underline{X}| = \frac{x_1 x_2 \gamma}{\sqrt{x_1^2 + x_2^2}} = \frac{x_1 x_2 \gamma}{|\underline{\xi}|} \qquad (B\text{-}3)$$

Using polar coordinates, $x_1 = |\underline{\xi}| \cos\theta$, $x_2 = |\underline{\xi}| \sin\theta$:

$$\underline{e} = |\underline{\xi}| \gamma \cos\theta \sin\theta \qquad (B\text{-}4)$$

Since in pure shear deformation dilatation is zero, we have $\underline{e} = \underline{e}^d$. For the yield function based on the 2D force state given in Eq. (49), with $f(\underline{t}^d) = 0$, on the yield surface:

$$\psi_0 = \frac{\|\underline{t}^d\|^2}{2} = \frac{1}{2}\left(\frac{8\mu}{m}\right)^2 \|\underline{e}^d\|^2 \qquad (B\text{-}5)$$

Where:

$$\|\underline{e}^d\|^2 = \int_H (\underline{e}^d)^2 \, dV_\xi \qquad (B\text{-}6)$$

Using cylindrical coordinate $dV_\xi = l_z |\underline{\xi}| \, d\xi d\theta$, where $l_z$ is the thickness of the system in the third direction in 2D condition to have volume with the same unit as classical and 3D case:

Then we have:

$$\|\underline{e}^d\|^2 = \int_H (\underline{e}^d)^2 \, dV_\xi = \int_0^{2\pi} \int_0^\delta |\underline{\xi}|^2 \gamma^2 \cos^2\theta \sin^2\theta \, l_z \, |\underline{\xi}| \, d\xi d\theta = \frac{\pi l_z \gamma^2 \delta^4}{16} \qquad \text{(B-7)}$$

$$m = \int_H r^2 \, dV_\xi = \int_0^{2\pi} \int_0^\delta r^2 \, l_z \, |\underline{\xi}| \, d\xi d\theta \qquad \text{(B-8)}$$
$$= \frac{\pi l_z \delta^4}{2}$$

Substituting (B-7) and (B-8) in (B-5) gives the yield value:

$$\psi_0 = \frac{8\tau_y^2}{\pi l_z \delta^4} = \frac{8}{3} \frac{\sigma_y^2}{\pi l_z \delta^4} \qquad \text{(B-9)}$$

Note that in the above equation we substituted $\mu \gamma_y$ with shear stress $\tau_y$ and then $\tau_y = \frac{\sigma_y}{\sqrt{3}}$ as explained in (Mitchell, 2011).

**Appendix C**

As it is explained in section 4.2.2, we modify our yield function (which is based on the deviatoric force state) for our 2D case by adding the "missing" out-of-plane contribution compare to classical von-Mises equivalent stress relation. The classical von-Mises stress for 2D plane tress /strain conditions is:

$$\sigma_{vm}^2 = \frac{3}{2} \sigma^d : \sigma^d = \frac{3}{2} \left( (\sigma_{11}^d)^2 + (\sigma_{22}^d)^2 + (\sigma_{33}^d)^2 + 2(\sigma_{12}^d)^2 \right) \qquad \text{(C-1)}$$

Using the yield function given in Eq. (49) and relation for $\psi_0$ (Eq. (50)), we expect that $\frac{3\pi l_z}{8} \frac{\|t^d\|^2}{2} = \sigma_{vm}^2$. Since our 2D $t^d$ ignors the out-of-plane component which is exist in the deviatoric stress tensor for 2D plane stress/strain conditions, in our 2D case, we get:

$$\frac{3\pi l_z}{8} \frac{\|t^d\|^2}{2} \cong \frac{3}{2} \left( (\sigma_{11}^d)^2 + (\sigma_{22}^d)^2 + 2(\sigma_{12}^d)^2 \right) \qquad \text{(C-2)}$$

Comparing left hand side of (C-1) with left hand side of (C-2), shows that our yield function will be modified by adding a term equivalent to the missing out of plane component, $\frac{3}{2}(\sigma_{33}^d)^2$.

Using properties of deviatoric stress tensor that its trace is zero, we have:

$$\sigma_{33}^d = -(\sigma_{11}^d + \sigma_{22}^d) \qquad \text{(C-3)}$$

We need to find $(\sigma_{11}^d + \sigma_{22}^d)$ in terms of our PD force state. Note that the PD collapsed deviatoric stress tensor we obtained for the 2D case has the following format:

$$\sigma_{2D-PD}^d = \begin{bmatrix} \sigma_{11}^d & 0 & 0 \\ 0 & \sigma_{22}^d & 0 \\ 0 & 0 & 0 \end{bmatrix} \tag{C-4}$$

Hence, $\sigma_{11}^d + \sigma_{22}^d$ is the trace of $\sigma_{2D-PD}^d$. Then, we tried to find a relationship that gives the trace of the PD stress tensor. For this reason, we used the following procedure:

The trace of PD isotropic stress tensor (see Eq. (19)) in the 2D case is equal to $2k\theta$, we can relate this term to its correspondence 2D isotropic force sate which is given in Eq.(27) as:

$$\text{trace}(\sigma_{2D-PD}^{iso}) = 2k\theta = \frac{\underline{t}^{iso} m}{\underline{\omega x}} \rightarrow \text{trace}(\sigma_{2D-PD}^{iso})\frac{\omega x}{m} = \underline{t}^{iso} \tag{C-5}$$

Dot product both side of the above equation with $\underline{x}$ results in:

$$\text{trace}(\sigma_{2D-PD}^{iso}) = \underline{t}^{iso} \bullet \underline{x} \tag{C-6}$$

Similar to the relation (C-6), the relation between the trace of PD deviatoric stress tensor and its correspondence deviatoric force state is:

$$\text{trace}(\sigma_{2D-PD}^d) = \underline{t}^d \bullet \underline{x} \rightarrow \sigma_{11}^d + \sigma_{22}^d = \underline{t}^d \bullet \underline{x} \tag{C-7}$$

Using relation (C-3) we have:

$$\sigma_{33}^d = -(\underline{t}^d \bullet \underline{x}) \tag{C-8}$$

now we have the out-of-plane component of deviatoric force state in terms of $\underline{t}^d$. Using (C-8), we modified the yield function as:

$$\frac{3\pi h \delta^4}{8}\frac{\|\underline{t}^d\|^2}{2} + \frac{3}{2}(\underline{t}^d \bullet \underline{x})^2 \cong \sigma_{vm}^2 \rightarrow$$
$$\psi(\underline{t}^d) = \frac{\|\underline{t}^d\|^2}{2} + \frac{4(\underline{t}^d \bullet \underline{x})^2}{\pi h \delta^4} \tag{C-9}$$

This equivalent form of von-Mises based on the deviatoric force sate is used in yield criteria given in Eq. (51).

**Appendix D**

The final deviatoric force state $(\underline{t}_{n+1}^d)$ in terms of $\underline{t}_{trial}^d$ and $\Delta\lambda$, can be obtained by updating the plastic part of deviatoric extension state:

For plane stress condition:

$$\underline{t}^d_{n+1} = \left(-k - \frac{8\mu}{3}\right)\theta^e \frac{\underline{\omega x}}{m} + \frac{8\mu}{m} \underline{\omega} \ \underline{e}^e_{n+1} \tag{D-1}$$

Using the relation for $\theta$ in-plane stress condition:

$$\underline{t}^d_{n+1} = \left(-k - \frac{8\mu}{3}\right)\frac{2(2\nu-1)}{\nu-1}\left(\frac{\underline{\omega x}}{m} \bullet \underline{e}^e_{n+1}\right)\frac{\underline{\omega x}}{m} + \frac{8\mu}{m}\underline{\omega}\ \underline{e}^e_{n+1}$$
$$= \left(-k - \frac{8\mu}{3}\right)\frac{2(2\nu-1)}{\nu-1}\left(\frac{\underline{\omega x}}{m} \bullet (\underline{e}_{n+1} - \underline{e}^{dp}_n - \Delta\lambda\nabla^d\psi)\right)\frac{\underline{\omega x}}{m}$$
$$+ \frac{8\mu}{m}\underline{\omega}\ (\underline{e}_{n+1} - \underline{e}^{dp}_n - \Delta\lambda\nabla^d\psi)$$

Using Eq. (58):

$$\underline{t}^d_{n+1} = \underline{t}^d_{trial} - \frac{8\mu}{m}\underline{\omega}(\Delta\lambda\nabla^d\psi) + \frac{2(2\nu-1)}{\nu-1}\left(k + \frac{8\mu}{3}\right)\frac{\underline{\omega x}}{m}\left(\frac{\underline{\omega x}}{m} \bullet \Delta\lambda\nabla^d\psi\right) \tag{D-2}$$

Repeating the same procedure to define $\underline{t}^d_{n+1}$ for plane strain condition result in a relation proposed in Eq. (62). For obtaining the $\nabla^d\psi$ in that equation, we have $\psi(\underline{t}^d) = \frac{\|\underline{t}^d\|^2}{2} + \frac{4(\underline{t}^d \bullet \underline{x})^2}{\pi h \delta^4}$ in which $\underline{t}^d \bullet \underline{x}$ obtained by using Eq. (28):

$$\underline{t}^d \bullet \underline{x} = \begin{cases} \left(-k - \frac{8\mu}{3}\right)\theta + \frac{8\mu}{m}\underline{\omega e} \bullet \underline{x} & \text{plane stress} \\ -\frac{10\mu}{3}\theta + \frac{8\mu}{m}\underline{\omega e} \bullet \underline{x} & \text{plane strain} \end{cases} \tag{D-3}$$

where:

$$\underline{\omega e} \bullet \frac{\underline{x}}{m} = \begin{cases} \frac{\theta(\nu-1)}{2(2\nu-1)} & \text{plane stress} \\ \frac{\theta}{2} & \text{plane strain} \end{cases} \tag{D-4}$$

Based on the relations (D-3) and (D-4), $\underline{t}^d \bullet \underline{x}$ is only in terms of $\theta$ for both plane stress and plane strain condition. Hence, it does not play any role in defining $\nabla^d\psi$ an the only effective term is $\frac{\|\underline{t}^d\|^2}{2}$. Therefore, we have:

$$\nabla^d\psi = \nabla^d\left(\frac{\|\underline{t}^d\|^2}{2}\right) = \underline{t}^d \tag{D-5}$$

**Appendix E:**

In this part, we want to show the consistency of the proposed yield functions with classical J2 plasticity. For this reason, the von-Mises equivalent stress is calculated using Eq. (53). The results are presented in Figure 14 (Note that $\delta = 2.5$ mm, $m = 5$):

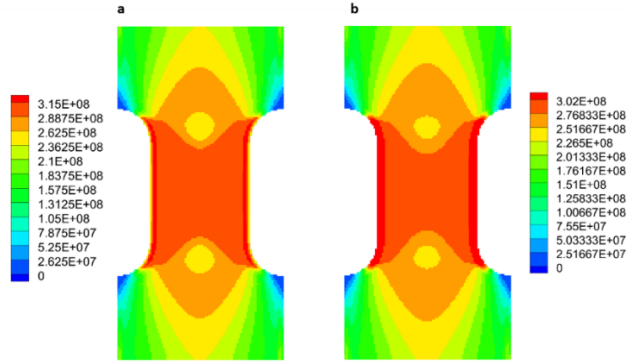

Figure 14: von-Mises equivalent stress (using Eq. (53)) distribution in PD with yield function based on deviatoric force state (a), and deviatoric strain energy density (b).

Although, the PD results for von-Mises stress distribution using Eq. (52) for two different yield functions (**Error! Reference source not found.**) are close to each other, when we use Eq. (53) to compute von-Mises stress, the obtained results with two yield functions are different in terms of predicting the yield value $\sigma_y = 300$ MPa. Using the deviatoric strain energy-based yield function leads to a yield value closer to the targeted one than using the yield function based on the deviatoric force state. Also notice the presence of the peridynamic surface effect near the left and right boundaries. This is expected, since only when a peridynamic node has a full neighborhood region, the collapsed stress tensor (used in the calculation of $\sigma_{vm}$ Eq. (53) ) converges to classical results. The thickness of the surface layer will decrease by using the smaller horizon size. This is seen from the results in Figure 15:

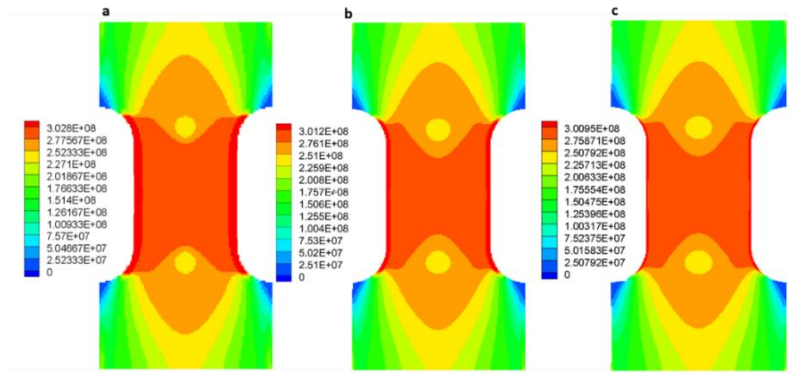

Figure 15: The PD von-Mises stress distribution (using Eq. (53)) for the yield function based on the deviatoric strain energy density with *m*=5 and the horizon sizes: 2.5mm (a), 1.25mm (b) and 0.625 mm (c).